\documentclass[12pt]{iopart}
\pdfoutput=1
\usepackage{graphicx}
\usepackage{dcolumn}
\usepackage{bm}
\usepackage{tikz}
\usepackage{amssymb, setstack}

\begin{document}
\title[The effect of anisotropy on pedestrian flows]{The effect of perception anisotropy on particle systems describing pedestrian flows in corridors}

\author{Lennart Gulikers$^1$, Joep Evers$^2$, Adrian Muntean$^2$ and Alexey Lyulin$^3$}
\address{$^1$ Department of Mathematics and Computer Science, Eindhoven University of Technology, P.O. Box 513, 5600 MB Eindhoven, The Netherlands}
\address{$^2$ Centre for Analysis, Scientific computing and Applications, Department of Mathematics and Computer Science \& Institute for Complex Molecular Systems, Eindhoven University of Technology, P.O. Box 513, 5600 MB Eindhoven, The Netherlands}
\address{$^3$ Theory of Polymers and Soft Matter, Department of Applied Physics, Eindhoven University of Technology, P.O. Box 513, 5600 MB Eindhoven, The Netherlands}
\eads{\mailto{l.gulikers@student.tue.nl}, \mailto{j.h.m.evers@tue.nl} (corresponding author), \mailto{a.muntean@tue.nl}, \mailto{a.v.lyulin@tue.nl}}

\begin{abstract}
We consider a microscopic model (a system of self-propelled particles) to study the behaviour of a large group of pedestrians walking in a corridor. Our point of interest is the effect of anisotropic interactions on the global behaviour of the crowd. The anisotropy we have in mind reflects the fact that people do not perceive (i.e. see, hear, feel or smell) their environment equally well in all directions. The dynamics of the individuals in our model follow from a system of Newton-like equations in the overdamped limit. The instantaneous velocity is modelled in such a way that it accounts for the angle under which an individual perceives another individual.\\
We investigate the effects of this perception anisotropy by means of simulations, very much in the spirit of molecular dynamics. We define a number of characteristic quantifiers (including the polarization index and Morisita index) that serve as measures for e.g. organization and clustering, and we use these indices to investigate the influence of anisotropy on the global behaviour of the crowd. The goal of the paper is to investigate the potentiality of this model; extensive statistical analysis of simulation data, or reproducing any specific real-life situation are beyond its scope.
\end{abstract}
\noindent{\it Keywords\/}: Traffic and crowd dynamics, interacting agent models, self-propelled particles, pattern formation (Theory)

\pacs{05.10.Ln, 05.20.-y, 05.60.-k, 05.90.+m, 89.75.Fb}
\ams{82Cxx, 65Pxx, 65Zxx, 37Nxx, 37Fxx}


\section{\label{sec:level1}Introduction}
During the last two, three decades, the field of crowd dynamics has emerged as the \textit{natural} sciences' reaction to questions arising from \textit{social} sciences, population biology and urban planning. See e.g.~\cite{Stott} for an example of a problem addressed in psychology, or \cite{Carroll} for an illustration of the civil engineering aspects. The roots and philosophy of crowd dynamics are very much in the spirit of statistical mechanics, molecular dynamics, interacting particle systems methods and the theory of granular matter, as such treating individual humans nearly as non-living material (cf.~e.g.~the nice overview \cite{Schadschneider2011} and references cited therein). A justification for this approach lies in the fact that the individuals' personal will is more or less averaged out if one looks at the crowd as a whole. From this perspective, it can be considered as (stochastic) noise, superimposed on some `clean' (deterministic) dynamics.\\
To illustrate the thin borderlines between several fields of study, the reader is referred to e.g.~\cite{Lutz2,Lutz1} for the dynamics of non-living particles, \cite{Lega,Peruani} for studies of tumbling or self-propelled living particles (like bacteria), or \cite{Molnar, CRAS, CristianiTosin} for crowd dynamics. Although these fields all focus their own specific real-world scenario, their way of thinking and posed questions are very much alike.\\
\\
However, an evident and important difference between people and molecules or grains (apart from people's own opinions, irritations etc.) is the fact that people clearly have front and back sides. Our degree of perceiving our surroundings highly depends on the direction of looking. We mainly base our walking behaviour on what we see, and clearly what happens in front of us thus has more influence than what happens behind us (this statement is also supported e.g.~by \cite{Guo12}). A modification or extension of physics-inspired models is needed to incorporate this kind of anisotropy in the interactions between individuals. This paper investigates the effect of anisotropy on the global behaviour of a group of pedestrians.\\
\\
Our focus is  on the simulation of a scenario where pedestrians move in a long corridor. We might relate this situation to evacuation of people from a building (cf. \cite{Armin, HelbingNature}). It is sane to assume that these evacuees have an intrinsic \textit{drive} to move towards the exit (i.e. one side of the corridor), and moreover that there view is focused in the same direction. Investigating the effect of anisotropy on the large-scale behaviour of the crowd therefore relates to assessing the escape process.\\
\\
In Section \ref{sec:level1} of this paper the model is presented and explained. Section \ref{sec:sim setup and results} is the main part of the paper. It describes the exact scenario of our simulations and the definitions of the quantities we use for assessing the results (polarization index, projected density, Morisita index). Moreover, in this section the simulation results are presented and discussed. Conclusions and an outlook on possible future work are given in Section \ref{sec:conclusion}.

\section{\label{sec:level1}A model for anisotropic interactions between pedestrians}
We represent pedestrians  by point particles\footnote{Note that this is a modelling choice, and that one can view crowds from different perspectives, e.g.~micro-, meso- and macroscopic, or a suitable combination of these. See \cite{Bellomo} for a critical review, or the broader overview in \cite{Schadschneider2011}.} having masses $m_i$. They are located in a long corridor of length $\tilde{L}$ and width $B$. Here, the word `long' refers to the fact that at the time scales we focus on, the pedestrians are not able to reach the end of the corridor. Interactions between pedestrians are short-ranged. We therefore suppose that the correlation length in the system is less than or equal to a certain $L\ll\tilde{L}$ and we can subdivide the corridor in an array of rectangles (width $B$ and length $L$), which are all duplicates of each other. We thus have a scenario with periodic boundary conditions. Our domain of interest is therefore a rectangular box
\begin{equation}
\Omega:= [ -\frac{L}{2} , \frac{L}{2} ] \times [ -\frac{B}{2} , \frac{B}{2} ],
\end{equation} with periodic boundary conditions in one direction and impermeable walls in the other direction. The corridor contains $N < \infty$ pedestrians. For all $i$ $\in \{1, ... , N\}$ and $t\geq 0$, the vector $\vec{r}_i (t) = (x_i(t) , y_i(t)) \in \Omega$ represents the position of the $i$-th pedestrian at time $t$. We denote its velocity by $\vec{v}_i(t)$.\\
\\
We assume that the governing equation of motion is
\begin{equation}
\frac{m_i}{\tau_{drive}} (\vec{v}_i(t) - \vec{v}_{des}) =  \vec{F}_i^{soc} + \vec{F}_i^{phys}.
\label{newton_overdamped}
\end{equation}
The equation describes the motion of the $i$-th individual, which has mass $m_i$ and which moves with velocity $\vec{v}_i(t)$. However, he/she \textit{tries} to move according to its desired velocity $\vec{v}_{des}$. Here, $\tau_{drive}$ is the characteristic relaxation time related to attaining the desired velocity. Its actual velocity is moreover perturbed by two `forces'. The word `force' is used since (\ref{newton_overdamped}) can be regarded as an overdamped limit of a Newton-like equation (cf.~\cite{Molnar} for this Newton-like way of modelling).\\
One could argue whether the social force is the right concept to use to drive the pedestrians, or maybe ideas like social pressure (as in a Darcy-like law) or cognitive-based heuristics (see e.g. \cite{Moussaid}) are more appropriate.  Here we avoid any polemic by deciding to choose a framework based on social forces and leave for later any further developments of other possible approaches.\\
\\
There is a physical force $\vec{F}_i^{phys}$ that acts on the individual to describe the effect of the non-living environment (geometry). In this paper we only take into account the influence of walls on pedestrians, that is: $\vec{F}_i^{phys}$ = $\vec{F}_i^{wall}$. Furthermore, pedestrian $i$ experiences a so-called social force $\vec{F}_i^{soc}$ due to the presence of other individuals, which influences the motion of this particular pedestrian $i$.\\
\\
Individuals are influenced by the walls as soon as they come too close, i.e. within a distance $R_{wall}$. We model these impermeable walls by means of a strong repulsive force $\vec{F}_i^{wall}$ acting on pedestrian $i$:
\begin{equation}
\vec{F}_i^{wall} = \left\{
                     \begin{array}{ll}
                       F_{Wall}\, (1 - \frac{R_{wall}}{d}) \vec{n}, & \hbox{if $d<R_{wall}$;} \\
                       \vec{0}, & \hbox{otherwise.}
                     \end{array}
                   \right.
\label{FRight}
\end{equation}
Here, $\vec{n}$ is the unit normal pointing from the corresponding wall into the corridor, $F_{Wall}$ is the strength of the repulsive force and $d$ is the distance to the wall for pedestrian $i$. The word `strong' here implies that this force is not just a contact force, but has a longer range. Typically, this makes individuals avoid walls before touching them.\footnote{We will see later that the type of repulsion in $\vec{F}_i^{wall}$ for small $d$ is the same as in the interactions between individuals (cf.~(\ref{Fsoc})--(\ref{U_attr_rep})), be it with different parameters.}\\
\\
Furthermore, very much in the spirit of \cite{Molnar}, we specify the social force by
\begin{equation}
\vec{F}_i^{soc} = \sum_{\vec{r}_j \in \Omega_i} - \nabla W(\vec{r}_i - \vec{r}_j),
\label{Fsoc}
\end{equation}
where:
\begin{itemize}
\item $\Omega_i$ is the collection of the position vectors of all individuals which are within a distance $R_{cut}$ to pedestrian $i$. In other words, pedestrians interact only when they are close enough to each other;
\item we assume that the interaction potential $W$ depends only on the relative position of the two pedestrians $i$ and $j$ and not on their relative velocity.
\end{itemize}

\begin{figure}[ht]
\centering
\begin{tikzpicture}[scale=1.0, >= latex]
	\draw[->] (0,0)--(0,1)node[anchor=north east]{$\vec{e}_y$};
	\draw[->] (0,0)--(1,0)node[anchor=north east]{$\vec{e}_x$};

    \draw[->, thick] (0,0)--(1,2)node[anchor=north east]{$\vec{r}_i$ \hspace{0.5 cm}};
    \draw[->, thick] (0,0)--(3,4)node[anchor=north east]{ \hspace{0.5 cm}};

    \draw[->, thick] (1,2)--(3,4) node[midway, sloped, above]{$\vec{r}_j-\vec{r}_i$} node[anchor=north west]{$\vec{r}_j$};

    \draw[->, thick] (1,2)--(3.5,2)  node[midway, sloped, below]{$\vec{v}_{des}$};

    \draw[->] (2.0, 2) arc (0:45:1.0);
    \node at (2.25,2.5) {$\theta_{r_i r_j}$};
\end{tikzpicture}
\caption{Schematic drawing of the variables $\vec{r}_i$, $\vec{r}_j$, $\vec{r}_j-\vec{r}_i$, involved angle $\theta_{r_i r_j}$ and unit vectors $\vec{e}_x$, $\vec{e}_y$. The vector $\vec{v}_{des}$ can in principle have arbitrary direction. For the sake of clarity in the picture, it is chosen to be parallel to $\vec{e}_x$. Here, $\theta_{r_i r_j}$ is the angle under which an individual positioned in $\vec{r}_i$ and heading along $\vec{v}_{des}$, perceives location $\vec{r}_j$.}\label{figure geometry}
\end{figure}
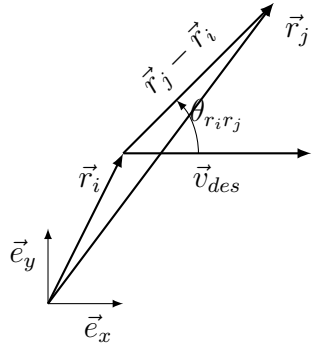

\noindent Specifically, $W$ takes the form,
\begin{equation}\label{W}
W(\vec{r}_i - \vec{r}_j) = U(|\vec{r}_i - \vec{r}_j|)\, \left( 1-\frac{\alpha}{2}(1-\cos\theta_{r_i r_j})\right ).
\end{equation}
Here, $\alpha \in [0,1]$ is an \textit{anisotropy parameter}, for which $\alpha=0$ means that the potential is isotropic, and $\alpha=1$ means that the anisotropy effects are maximal. The angle of perception $\theta_{r_i r_j}$ is the angle under which an individual positioned in $\vec{r}_i$, while moving in the direction of $\vec{v}_{des}$, perceives location $\vec{r}_j$, see Figure \ref{figure geometry}. The precise type of interaction is hidden in the structure of the function $U$. In particular, we distinguish between two types of interactions, namely
\begin{itemize}
\item \textit{only repulsive interaction} (for simplicity denoted by \underline{R case}). Pedestrians repel each other when their separation distance is smaller than $R_r^R$ (called radius of repulsion in \underline{R case}) and do not interact at larger distances;
\item \textit{both repulsive and attractive interaction} (denoted by \underline{AR case}). For this type of interaction, individuals repel each other when their separation distance is smaller than $R_r^{AR}$ (called repulsive radius in \underline{AR case}). However, when they are separated by a distance between $R_r^{AR}$ and $R_a^{AR}$ (called attractive radius in AR case), they are attracted to one another. They do not interact outside these regions.
\end{itemize}
For these two cases $R_{cut}=R_r^R$, respectively $R_{cut}=R_a^{AR}$. The second case is an extension of the first one. The \underline{R case} we might regard as a population of individualistic people that simply try to avoid each other. In the \underline{AR case} there is also some \textit{social cohesion}, as they try to keep the group together.
In the following, we describe the precise structure of our potentials.

\subsection{R case}
\begin{figure}[!ht]
\centering
\vspace{0cm}
\includegraphics[scale=0.5]{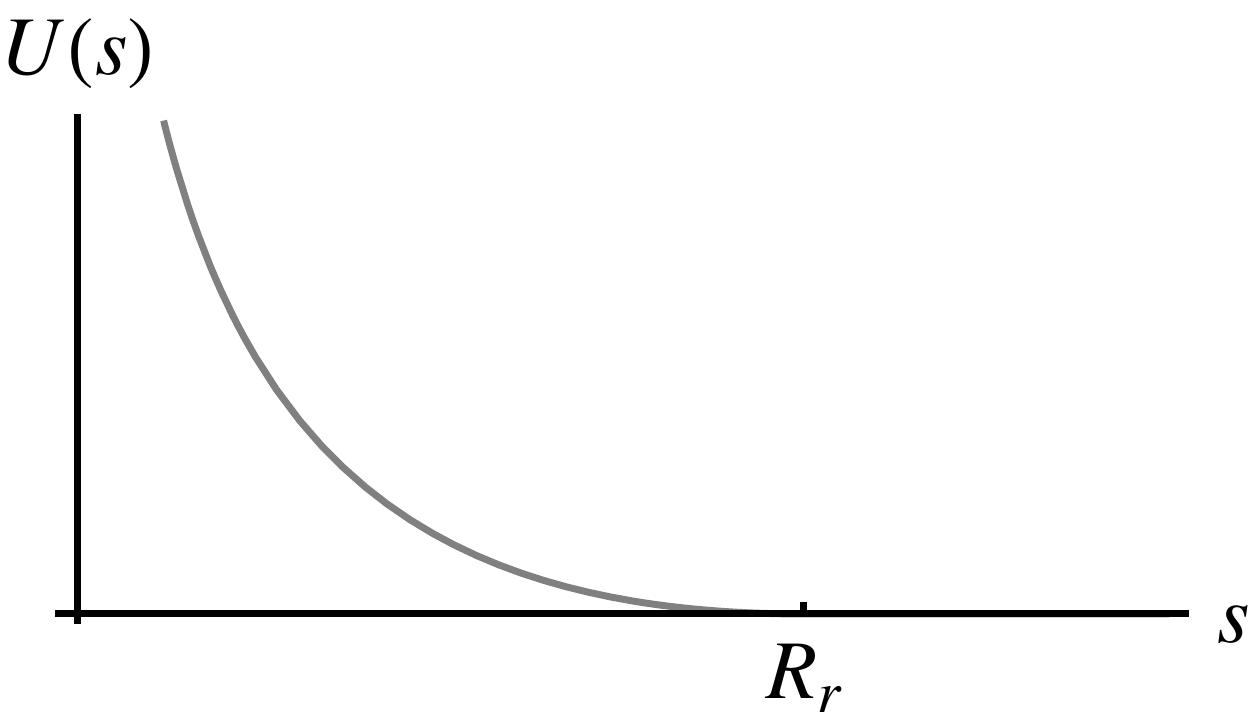}
\vspace{0cm}
\caption{Typical example of the interaction potential as given in (\ref{U repulsive}).}
\label{prim_r}
\end{figure}

In the repulsive case we take
\begin{equation}\label{U repulsive}
U(s) = \left\{
                \begin{array}{ll}
                  F^R(s-R^R_r-R^R_r\ln\frac{s}{R^R_r}) & \hbox{if $s<R^R_r$;} \\
                  0 & \hbox{if $s>R^R_r$.}
                \end{array}
              \right.
\end{equation}
See Figure \ref{prim_r} for an example of an interaction potential of the above form.
\subsection{AR case}
\begin{figure}[!ht]
\centering
\vspace{0cm}
\includegraphics[scale=0.5]{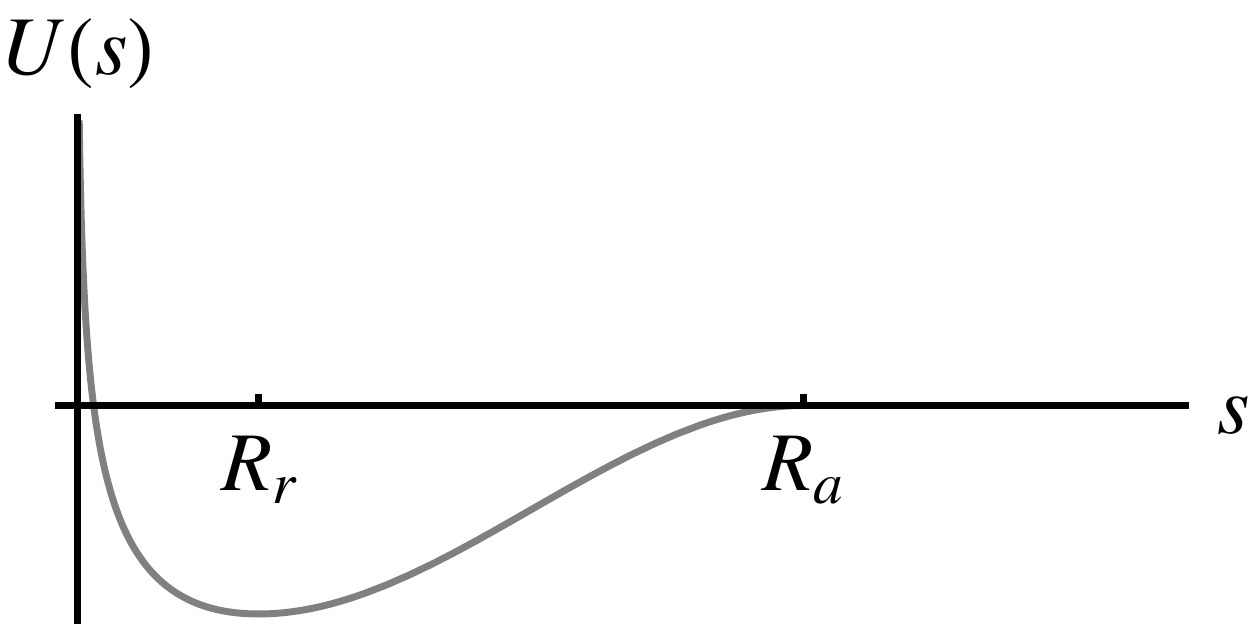}
\vspace{0cm}
\caption{Typical example of the interaction potential as given in (\ref{U_attr_rep})--(\ref{U attr-rep parabola}).}
\label{prim_ar}
\end{figure}

If we want to involve a both attractive and repulsive way of interactions, then we take
\begin{equation}
U(s) = \left\{
                \begin{array}{ll}
                  F^{AR}(s-R^{AR}_r\ln\frac{s}{R^{AR}_r})+C_2 & \hbox{if $s<R^{AR}_r$;} \\
                  \tilde{U}(s) & \hbox{if $R^{AR}_r<s<R^{AR}_a$;} \\
                  0 & \hbox{if $s>R^{AR}_a$;}
                \end{array}
              \right.
\label{U_attr_rep}
\end{equation}
where
\begin{equation}\label{U attr-rep parabola}
\tilde{U}(s)=\frac{F^{AR}(\frac{s^3}{3}-(R^{AR}_r+R^{AR}_a)s^2+R^{AR}_rR^{AR}_as)}{R^{AR}_r(R^{AR}_a-R^{AR}_r)}+C_1.
\end{equation}
In the above, the constants $C_1$ and $C_2$ are such that $U$ is a continuous function. Figure \ref{prim_ar} shows an example of this kind of interaction potentials.

\section{\label{sec:sim setup and results}Simulation: set-up and results}
\begin{figure}[!ht]
\centering
\vspace{-2cm}
\includegraphics[scale=0.5]{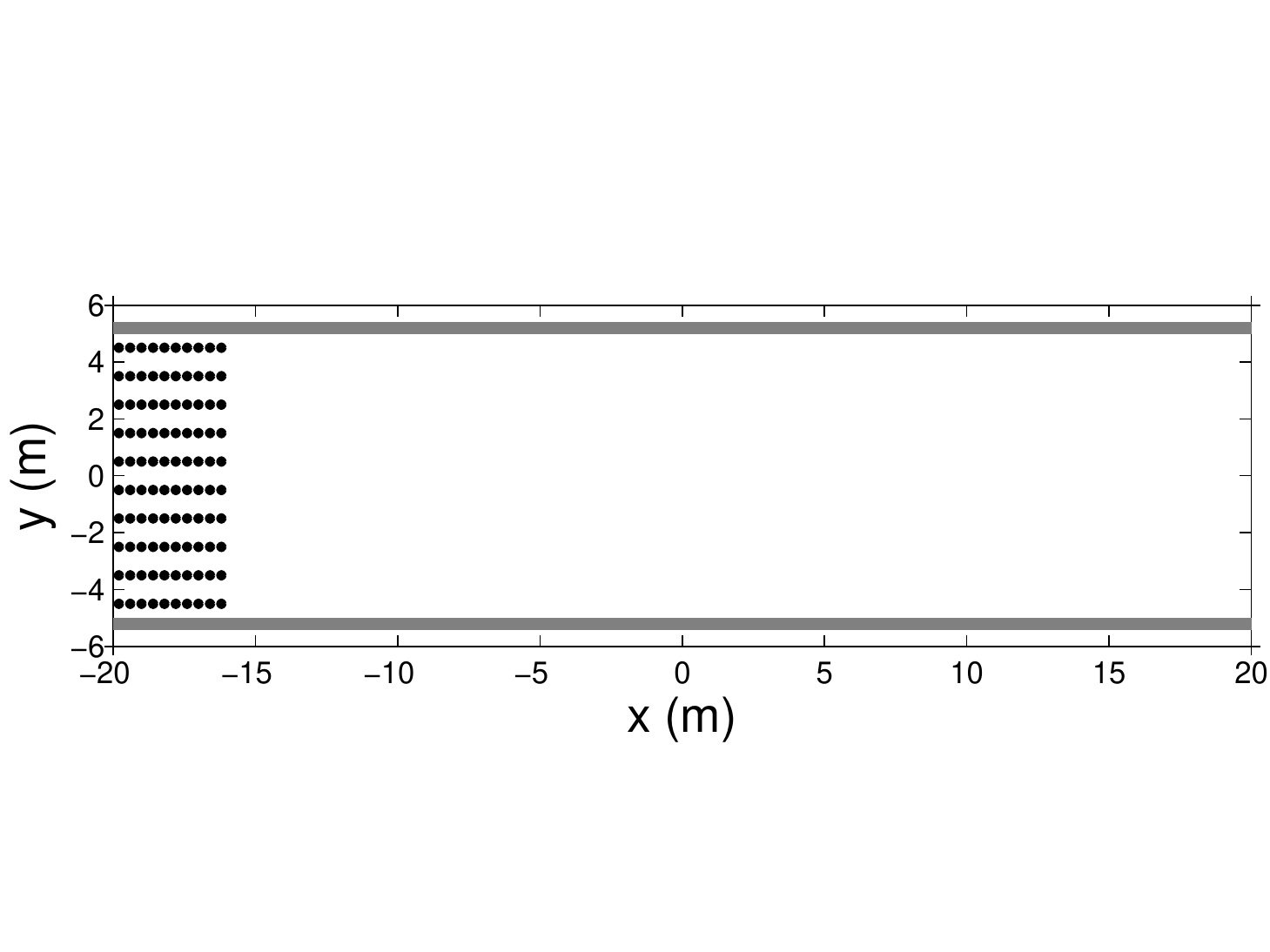}
\vspace{-1.5cm}
\caption{A snapshot of the positions of the 100 pedestrians at time $t=0$ s. This initial configuration is used in all simulations. The markers that indicate the individuals' positions are smaller than in Figures \ref{pos_r_00}--\ref{pos_ar_10}, just to avoid overlap here.}
\label{pos_0}
\end{figure}

For all simulations, the number of pedestrians $N$ is an integer multiple of 10. Initially, we place these individuals on a lattice of $N_x = \frac{N}{10}$ rows of 10 pedestrians each. The distance between two succeeding rows is always $\Delta x = \frac{L}{N}$. Note that there is an $N$ here, not $N_x$: the pedestrians are thus distributed initially in a domain about one tenth the length of the total corridor (see Figure \ref{pos_0}). The distance between two pedestrians in the same row is $\Delta y = \frac{B}{10}$. Let $(l_i,y_j)$ denote the position of pedestrian $k:= i + 10(j-1)$, where $i=1,\ldots,10$ and $j=1,\ldots,N_x$. Then the coordinates are $l_i = \frac{-L}{2} + \Delta x (i - \frac{1}{2})$ and $y_j = \frac{-B}{2} + \Delta y (j - \frac{1}{2})$. We use these initial conditions for \textit{all} simulations.\\
\\

\noindent The simulation time corresponds to a real life time $\tau_{obs} = 100$ s. This time is large enough to witness a stable profile for all simulations. We give in Table \ref{modpar} a summary of the parameters and their values used in both the `repulsive' and `attractive and repulsive' interaction potentials.\footnote{Regarding the values in the table, we remark that one might be hesitant actually to call a domain of width $B=10$ m a \textit{corridor}. In our setting, the word \textit{corridor} refers to the ratio between length and width, rather than to the absolute values of $B$ and $L$. Note that this ratio of the `real' corridor is in fact even bigger than $L/B$, namely $\tilde{L}/B$, cf.~the explanation in the beginning of Section \ref{sec:level1}.}\\

\begin{figure}[!ht]
\centering
\vspace{-2cm}
\includegraphics[scale=0.5]{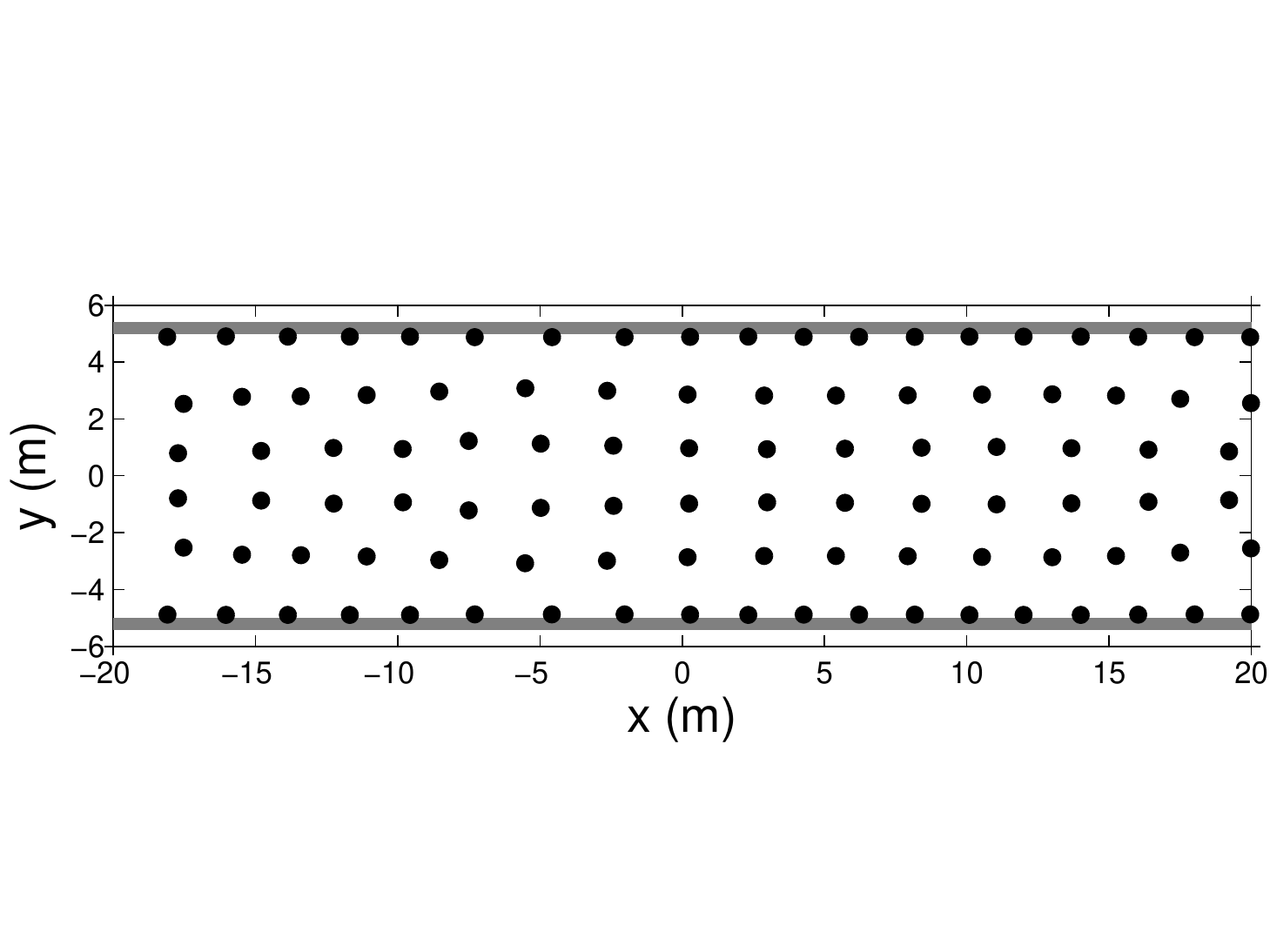}
\vspace{-1.5cm}
\caption{A snapshot of the positions of the 100 pedestrians at time $t = 100$ s, plot for the R potential with $\alpha = 1.0$ (most anisotropic interactions).}
\label{pos_r_00}
\end{figure}

\begin{figure}[!ht]
\centering
\vspace{-2cm}
\includegraphics[scale=0.5]{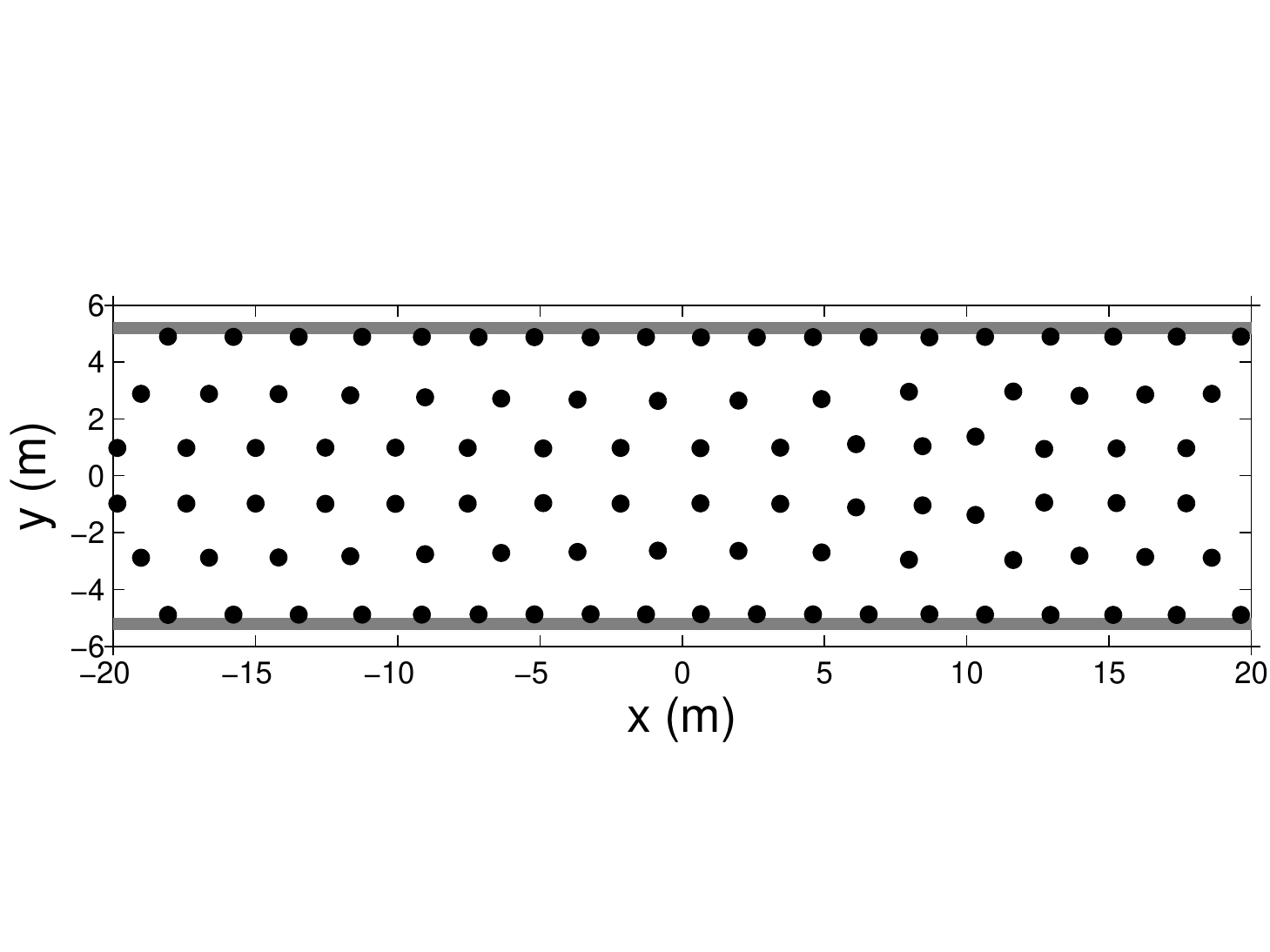}
\vspace{-1.5cm}
\caption{A snapshot of the positions of the 100 pedestrians at time $t = 100$ s, plot for the R potential with $\alpha = 0.5$.}
\label{pos_r_05}
\end{figure}

\begin{figure}[!ht]
\centering
\vspace{-2cm}
\includegraphics[scale=0.5]{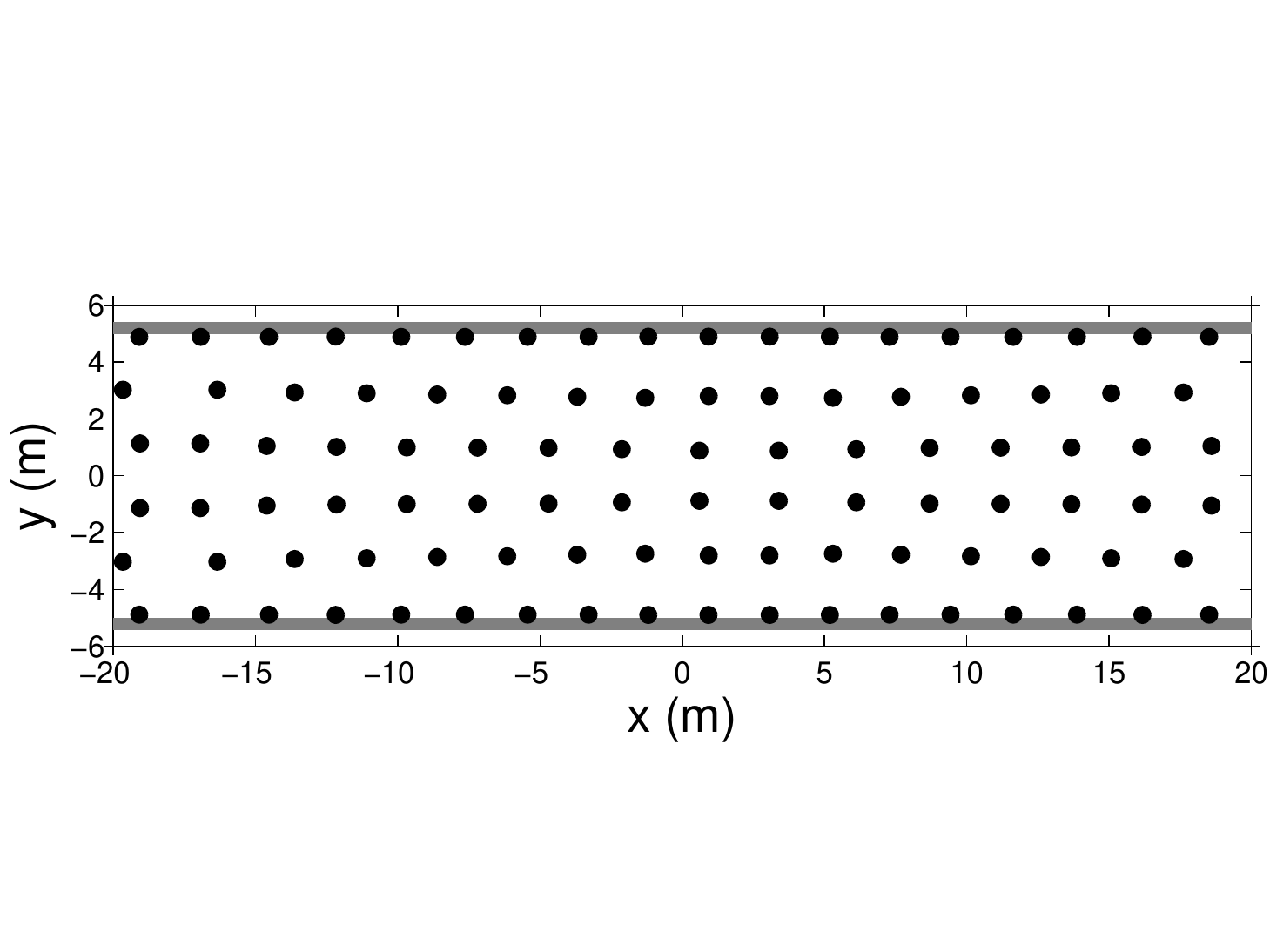}
\vspace{-1.5cm}
\caption{A snapshot of the positions of the 100 pedestrians at time $t = 100$ s, plot for the R potential with $\alpha = 0.0$ (isotropic interactions).}
\label{pos_r_10}
\end{figure}

\begin{figure}[!ht]
\centering
\vspace{-2cm}
\includegraphics[scale=0.5]{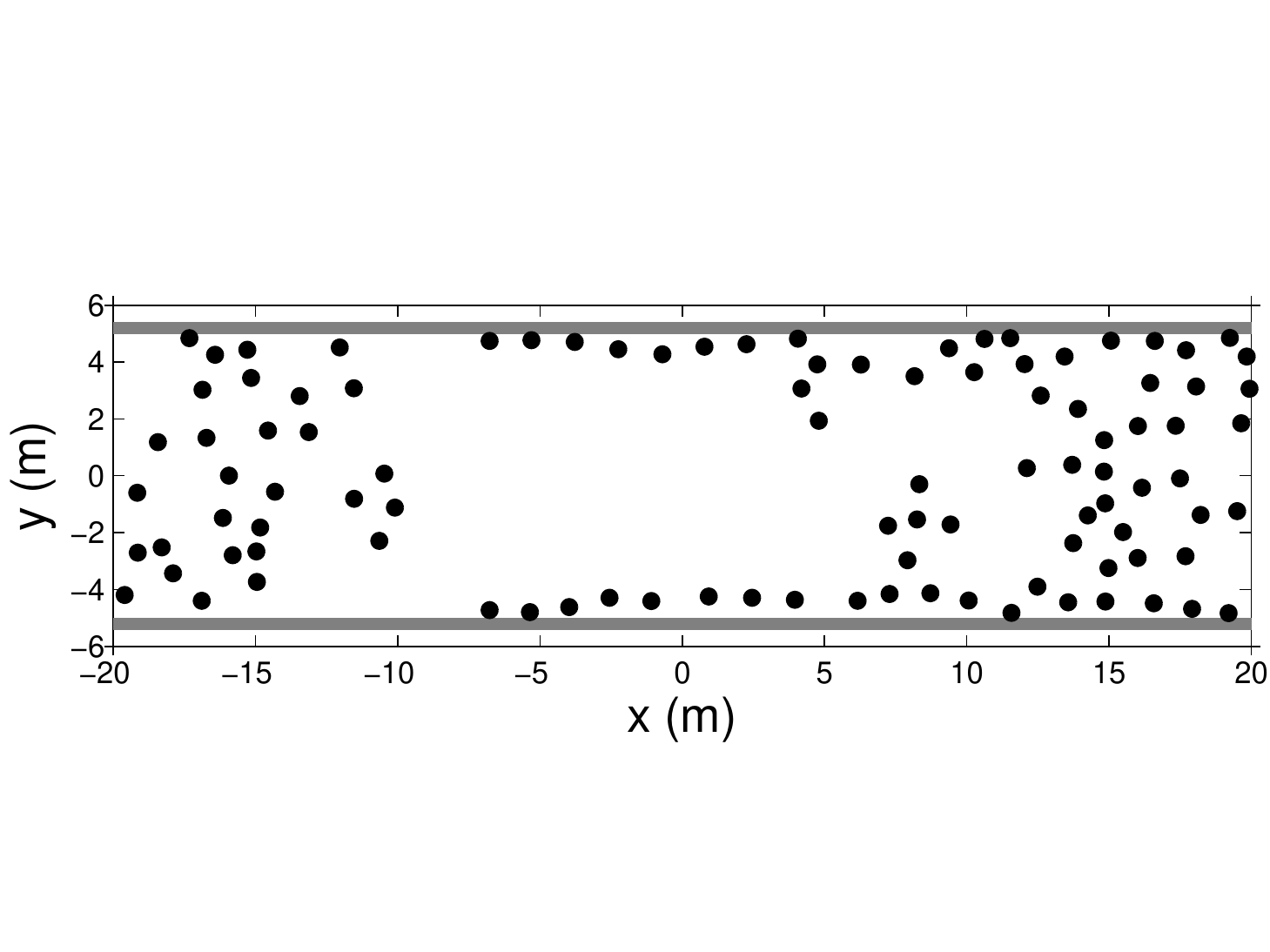}
\vspace{-1.5cm}
\caption{A snapshot of the positions of the 100 pedestrians at time $t = 100$ s, plot for the AR potential with $\alpha = 1.0$ (most anisotropic interactions). Note that the symmetry is broken due to discretization errors.}
\label{pos_ar_00}
\end{figure}

\begin{figure}[!ht]
\centering
\vspace{-2cm}
\includegraphics[scale=0.5]{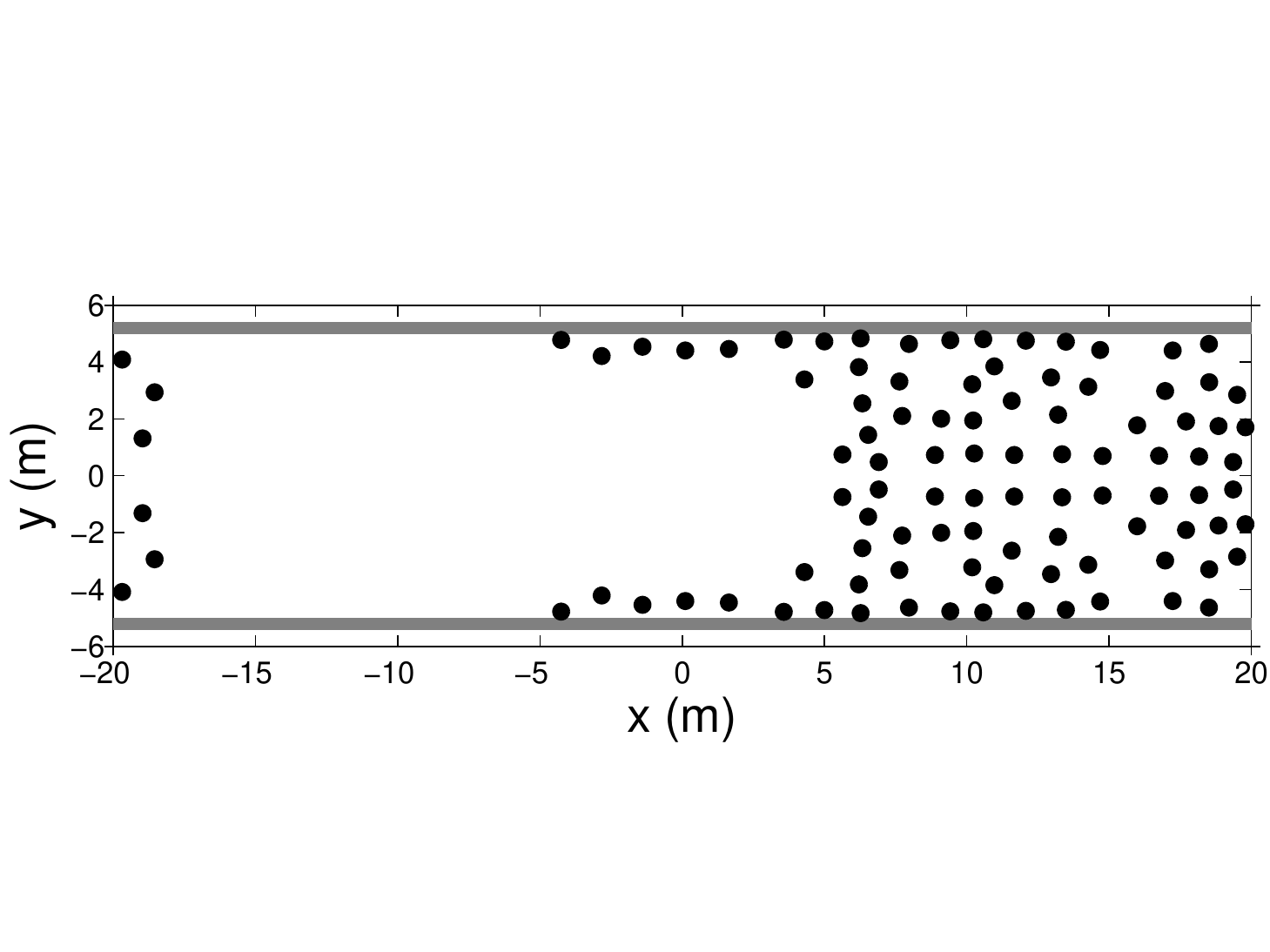}
\vspace{-1.5cm}
\caption{A snapshot of the positions of the 100 pedestrians at time $t = 100$ s, plot for the AR potential with $\alpha = 0.5$.}
\label{pos_ar_05}
\end{figure}

\begin{figure}[!ht]
\centering
\vspace{-2cm}
\includegraphics[scale=0.5]{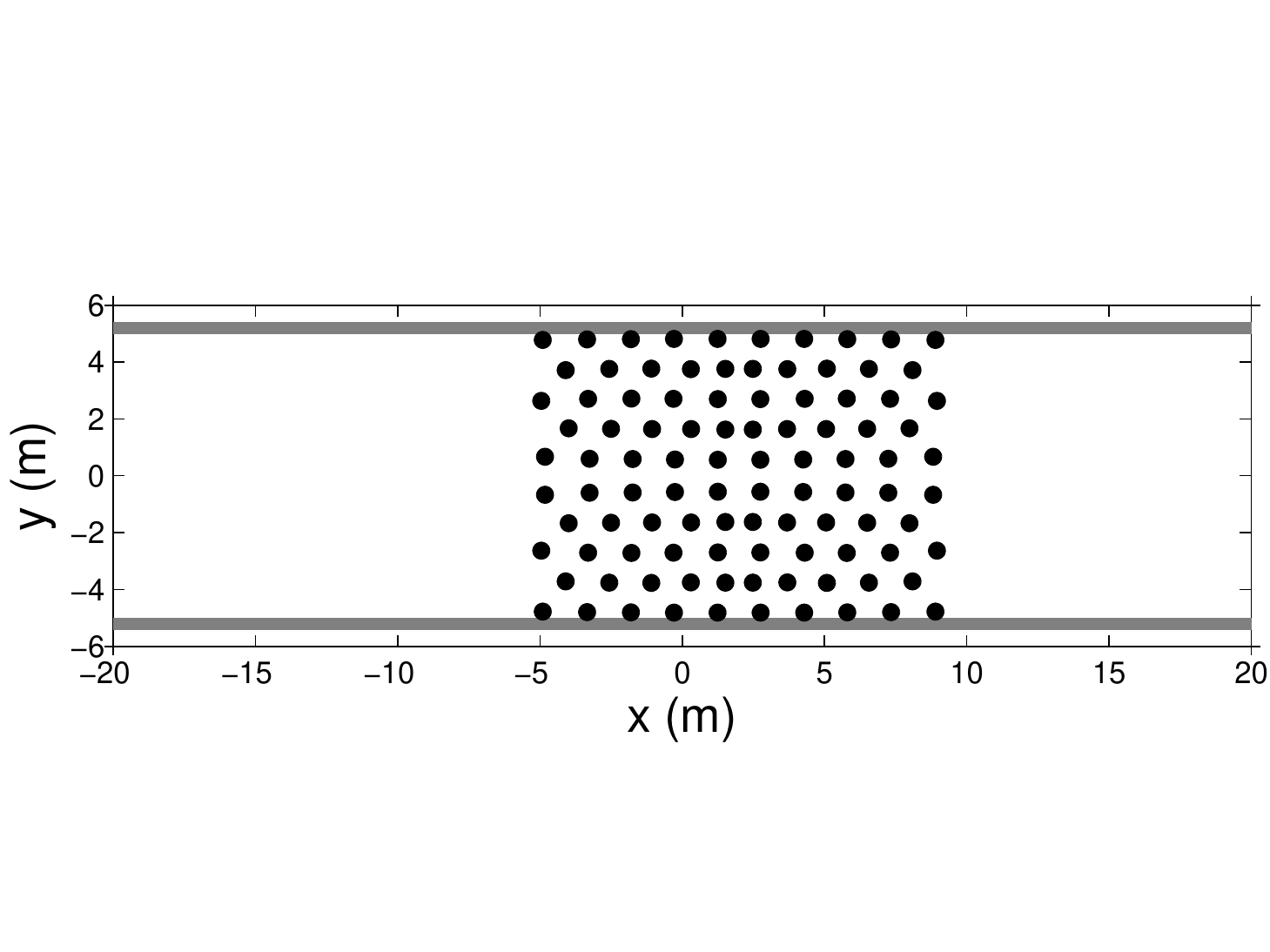}
\vspace{-1.5cm}
\caption{A snapshot of the positions of the 100 pedestrians at time $t = 100$ s, plot for the AR potential with $\alpha = 0.0$ (isotropic interactions). See main text for the used initial conditions.}
\label{pos_ar_10}
\end{figure}

\begin{table}[h]
\caption{Model parameters, symbols, units and reference values. We do simulations in two cases. The \underline{AR case}: pedestrians interact in a both attractive and repulsive way and the \underline{R case}: pedestrians only interact in a repulsive way. NB: $\mathcal{M}$  and $S_x \cdot S_y$ relate to the Morisita index and are specified in Section \ref{sect:def Morisita}.\\}\label{modpar}
\centering
\begin{tabular}{lcr}
\textrm{Symbol}&
\textrm{Unit}&
\textrm{Reference value}\\
\hline
$N$&-&20, 40, 60, 80, 100\\
$\alpha$&-&0.0, 0.5, 1.0\\
$\Delta t$&s&$0.0001$\\
$\tau_{drive}$&s&$1.0$\\
$\tau_{obs}$&s&$100$\\
$m$&kg&$50$\\
$B$&m&10\\
$L$&m&40\\
$\mathcal{M}$&-&64\\
$S_x \cdot S_y$& m$^2$ &2.5$\cdot$2.5\\
$F_r^R$&N&15\\
$F^{AR}$&N&15\\
$R_r^R$&m&4.0\\
$R_r^{AR}$&m&1.5\\
$R_a^{AR}$&m&3.0\\
\end{tabular}
\end{table}
\noindent For more technical details about this kind of simulations, the reader is referred e.g. to \cite{Ascher,Bornemann}.\\
\\
To illustrate simulation results, in Figures \ref{pos_r_00}--\ref{pos_ar_10} snapshots of the positions of $N=100$ pedestrians at time $t = 100$ s are shown, in both the \underline{R case} and the \underline{AR case} for $\alpha = 1.0 , \alpha = 0.5$ and $\alpha = 0.0$ (fully isotropic), respectively. Some individuals' positions might seem to coincide with the walls. Note however that our individuals are point particles and the walls have no width. The overlap is therefore only a result of the fact that we need to give size to individuals and walls in order to visualize them.

\noindent Except for Figure \ref{pos_ar_00}, all figures depict a completely symmetric profile with respect to the line $y = 0$. This is a natural result of the symmetry in the initial data and symmetry w.r.t. the direction of $\vec{v}_{des}$ in the interactions; it is just a property of the hyperbolic conservation laws to govern the mean-field behaviour, cf.~\cite{Dafermos}. Note also the strong tendency of the system to maintain (or produce) organized patterns. These issues will be addressed in more detail in Section \ref{sect: measured quant results}.\\
Figure \ref{pos_ar_00} is an exception in the sense that the distribution of the crowd is not symmetric around the line $y=0$. This cannot be explained from the model equations, as these imply that the distribution \textit{should} be symmetric. This effect must be due to round-off errors and their propagation with respect to time.\\
In the \underline{AR case} for $\alpha = 0.0$ (cf. Figure \ref{pos_ar_10},) we see that the people move in a crystal-like formation parallel to the $\vec{e}_x$ - axis, which remotely resembles the 2D crystallization patterns at low temperature pointed out in \cite{Theil}. The group as a whole is very compact. In the sequel we will use the words clusters/clustering for a situation in which pedestrians move close together.\\
However, if we compare Figure \ref{pos_ar_10} to Figures \ref{pos_ar_00} and \ref{pos_ar_05}, we observe that introducing anisotropy by setting $\alpha>0$ decreases the amount of structure and clustering.
In the \underline{R case}, all three values of $\alpha$ allow for a well-structured way of moving, but there is no clear clustering. It is evident from the figures that the people are spread over the whole corridor.\\
\\
Now, we introduce a number of measurable quantities to help quantify the above statements.

\subsection{\label{sec:level2}Definitions of the measured quantities}

\subsubsection{Polarization index}
Inspired by \cite{Lega}, we define the (time-dependent) polarization index $p$ of a group of people as the average angular deviation from the mean propagation direction. Here, the average is taken over all individuals. Note that zero polarization means that all the people move in parallel.\\
We first introduce the following auxiliary definitions:
\begin{itemize}
\item $\langle q \rangle_t := \frac{ \sum_{j = 1}^{M} q(t_j) } {M}$ denotes the time average of an arbitrary quantity $q$, based on the values $q(t_j)$ at time $t_j$, $j \in \{ 1,2, ... , M \}$;

\item $\theta_i(t) \in (-\pi , \pi]$ is the direction of motion of the pedestrian $i$ at time $t$. It is defined to be the angle $\theta_i(t)$ such that
\begin{equation}
\mbox{tan}(\theta_i(t))  = \frac{  \vec{v}_i(t) \cdot \vec{e}_y }{ \vec{v}_i(t) \cdot \vec{e}_x };
\end{equation}
\item $\theta(t) \in (-\pi , \pi]$ is the mean direction of motion of the pedestrians group. It is formally defined as the angle such that
\begin{equation}
\mbox{tan}(\theta(t))  = \frac{ \langle \vec{v}(t) \cdot \vec{e}_2 \rangle_N}{\langle \vec{v}(t) \cdot \vec{e}_1 \rangle_N},
\end{equation} where $\langle \vec{v}(t) \cdot \vec{e}_\xi \rangle_N$ denotes averaging over the total number of $N$
individuals:
\begin{equation}
\langle \vec{v}(t) \cdot \vec{e}_\xi \rangle_N := \frac{ \sum_{j = 1}^{N} \vec{v}_j(t) \cdot \vec{e}_\xi } {N},\, \xi \in \{ x,y \};
\end{equation}
\item $d\left(\theta_i(t),\theta(t)\right) := \min_{k\in\mathbb{Z}} |\theta_i(t)-\theta(t) + 2k\pi|$ denotes the angle in $[0 , \pi]$ between $\vec{v}_i(t)$ and the average direction of motion.
\end{itemize}
Now, the time-dependent polarization is defined as
\begin{equation}
p(t) := \frac{1}{N} \sum_{i = 1}^{N} d\left(\theta_i(t),\theta(t)\right).
\end{equation}
We are also interested in the time average of $p$, which is defined as
\begin{equation}
P := \left\langle p(t) \right\rangle_t = \left\langle \frac{1}{N} \sum_{i = 1}^{N} d\left(\theta_i(t),\theta(t)\right) \right\rangle_t.
\end{equation}

\subsubsection{Projection of the pedestrian density on the $\vec{e}_x$ axis}\label{sect:Sim, projection}
\begin{figure}[!ht]
\centering
\includegraphics[scale=0.5]{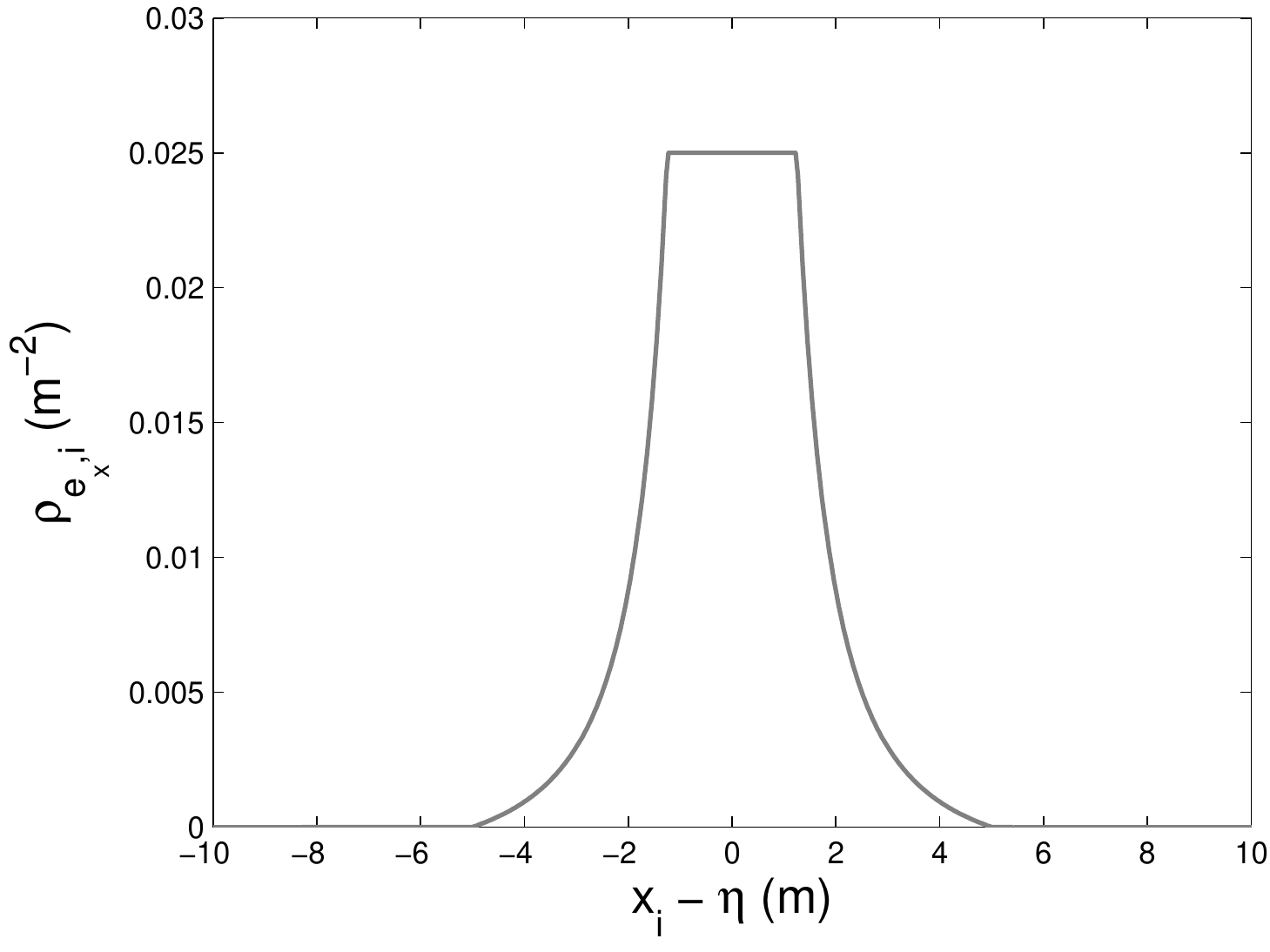}
\caption{The induced density on the $\vec{e}_x$ axis $\rho_{\vec{e}_{x,i}}$ prescribed by an individual as a function of the spatial coordinate $\eta$, relative to its horizontal coordinate $x_{i}$.}
\label{indDensity}
\end{figure}

To examine the distribution of our crowd in the direction that corresponds to the desired velocity, we can consider the number of pedestrians that are located in
\begin{equation}
S_{\eta}^{\epsilon} := \{ (x,y) \in \Omega : -\frac{B}{2} \leq y \leq \frac{B}{2}, |\eta - x|\leq \epsilon\}.
\end{equation}
Here, $0 < \epsilon \ll 1$, and $S_{\eta}^{\epsilon}$ denotes a narrow strip parallel to the $\vec{e}_y$ axis, centered at $x$-position $\eta$ and of width $\eta$. We could simply plot the number of people in $S_{\eta}^{\epsilon}$ as a function of $\eta$. Since pedestrians are point masses in our model, this procedure would produce a discontinuous, histogram-like graph. Moreover, the number of discontinuities would highly depend on the value of $\epsilon$. In order to smoothen the results, the individuals' coordinates have been projected on their corresponding $x$-coordinates. Next, we assign an `induced density' $\rho_{\vec{e}_{x,i}}$ to each individual $i$, which can be considered as a mollified Dirac delta distribution, see Figure \ref{indDensity}. It was constructed in such a way that it has support of width $\frac{L}{4}$.\\

\noindent The total density in a point $\eta \in [ - \frac{L}{2} , \frac{L}{2} ]$ is obtained by adding all individual contributions:
\begin{equation}
\rho_{\vec{e}_x}^N (\eta) := \sum_{i = 1}^{N} \rho_{\vec{e}_x,i} (\eta),
\end{equation} where the induced density $\rho_{\vec{e}_x,i}$ is given by,

\begin{equation}
\rho_{\vec{e}_x,i} (\eta) := \left\{
                \begin{array}{ll}
                   \frac{10}{BL} & \hbox{if $| x_i - \eta | \leq \frac{L}{32}$;} \\
                  \frac{L}{96 B | x_i - \eta |^2} - \frac{2}{3 BL} & \hbox{if $\frac{L}{32} \leq | x_i - \eta |  \leq \frac{L}{8}$;} \\
                  0 & \hbox{otherwise.}
                \end{array}
              \right.
\end{equation} Note that this density distribution is normalized:

\begin{equation}
\int_{\Omega} \rho_{\vec{e}_x}^N d\Omega = \sum_{i = 1}^{N} B \int_{-\frac{L}{2}}^{\frac{L}{2}} \rho_{\vec{e}_x,i} (x) dx = \sum_{i = 1}^{N} 1 = N.
\end{equation}

\subsubsection{Morisita index}\label{sect:def Morisita}
\begin{figure}[ht]
\centering
\begin{tikzpicture}[scale=0.8, >= latex]
	\draw[->] (6.5,0)--(6.5,1)node[anchor=north east]{$\vec{e}_y$};
	\draw[->] (6.5,0)--(7.5,0)node[anchor=north east]{$\vec{e}_x$};

    \fill[gray!10!white] (4,2) rectangle (-4,2.5);
    \draw[-, color=gray!30!white, line width = 0.5mm] (4,2)--(-4,2);
    \fill[gray!10!white] (4,-2) rectangle (-4,-2.5);
    \draw[-, color=gray!30!white, line width = 0.5mm] (4,-2)--(-4,-2);

    \draw[dashed] (4,2)--(4,-2);
    \draw[dashed] (3,2)--(3,-2);
    \draw[dashed] (2,2)--(2,-2);
    \draw[dashed] (1,2)--(1,-2);
    \draw[dashed] (-1,2)--(-1,-2);
    \draw[dashed] (-2,2)--(-2,-2);
    \draw[dashed] (-3,2)--(-3,-2);
    \draw[dashed] (-4,2)--(-4,-2);

	\draw[dashed] (4,-1.04)--(-4,-1.05);
	\draw[dashed] (4,1.05)--(-4,1.05);

    \shadedraw [ball color= black] (0.2,0.3) circle (0.1cm);
    \shadedraw [ball color= black] (0.5,0.5) circle (0.1cm);
    \shadedraw [ball color= black] (0.8,0.2) circle (0.1cm);

    \shadedraw [ball color= black] (1.2,1.6) circle (0.1cm);
    \shadedraw [ball color= black] (1.5,1.2) circle (0.1cm);

    \draw[dashed] (0,2)node[anchor=south]{$\frac{B}{2}$} --(0,-2)node[anchor=north]{$-\frac{B}{2}$};
	\draw[dashed] (4,0)node[anchor=west]{$\frac{L}{2}$}--(-4,0)node[anchor=east]{$-\frac{L}{2}$};

    \draw[<->] (-2,0)--(-2,-1)node[midway, right]{$S_y$};
    \draw[<->] (-2,0)--(-3,0)node[midway, anchor=south]{$S_x$};

\end{tikzpicture}
\caption{Illustration of the Morisita index. In this example, the corridor is subdivided into 32 boxes of dimensions $S_x \cdot S_y = \frac{L}{8} \cdot \frac{B}{4}.$ and 5 pedestrians are distributed over the boxes. This results in a Morisita index of $\frac{64}{5}$. }
\label{Mor}
\end{figure}
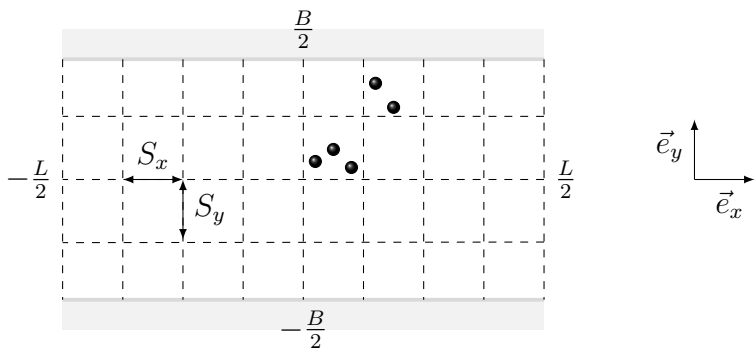

We subdivide the domain $\Omega$ in $\mathcal{M}$ equally sized rectangular boxes of dimensions $S_x \cdot S_y$. Inspired by \cite{Lega}, we define the Morisita index $I_{\mathcal{M}}$ for our system, which is a measure for the degree of dispersion (or of clustering) in our crowd. The index is the product of $\mathcal{M}$ and $\delta_\mathcal{M}$, where $\delta_\mathcal{M}$ is an estimator of Simpson's measure of diversity; cf.~\cite{Simpson}. This estimator is defined as
\begin{equation}
\delta_{\mathcal{M}} := \frac{ \sum_{i = 1}^{\mathcal{M}} n_i (n_i - 1) } { N (N-1) } ,
\end{equation}
where $n_i$ equals the number of pedestrians in box $i$. This is $\frac12\sum_{i = 1}^{\mathcal{M}} n_i (n_i - 1)$, the number of pairs consisting of two individuals within the same box, divided by $N (N-1)/2$: the total number of pairs in the system. Thus, $\delta_\mathcal{M}$ is an (unbiased) estimator of the probability that two randomly and independently chosen individuals are in the same box.\\
Now, assume that we draw two independent samples from a \textit{uniform} distribution on our corridor. The probability that both turn out to be in box $i$ is $(1/\mathcal{M})^2$, since they are independent and all boxes have the same size. Summation over all boxes leads to the probability that both sampled individuals are in the same, yet arbitrary, box: $\sum_{i = 1}^{\mathcal{M}} (1/\mathcal{M})^2 = 1/\mathcal{M}$.\\
\\
As mentioned above, the Morisita index, introduced in \cite{Morisita59, Morisita62}, is defined as
\begin{equation}\label{Morisita def}
I_{\mathcal{M}} := \mathcal{M}\, \delta_{\mathcal{M}}.
\end{equation}
The Morisita index can thus be interpreted as (an estimator of) the probability that two arbitrary pedestrians in our system are in the same box, divided by the probability that two uniformly distributed individuals are in the same box.\\
See Figure \ref{Mor} for an example that illustrates the use of the Morisita index. The figure also shows that the value of the Morisita index depends on the number of boxes $\mathcal{M}$ and their distribution.

\subsubsection{Connection between the Morisita index and entropy}\label{sect:Morisita Entropy}
We already mentioned the word `clustering'. In information theory, this concept refers to the process of ordering items: each \textit{document} should be assigned to a \textit{cluster}. The reader is referred to \cite{Manning}, Chapter 16, for more details. A measure to evaluate the quality of a clustering is the \textit{entropy}. Our aim here is to point out that this entropy is related to the Morisita index, be it that there is no one-to-one correspondence.\\
Let $K$ be a fixed integer number (the number of clusters) and define $\Omega=\{\omega_1,\ldots,\omega_K\}$ to be the set of clusters. One should see $\omega_k$ as the set of all documents assigned to cluster $k$. In total there are $N$ documents. We follow \cite{Manning}, pp.~328--329, in defining the entropy $H(\Omega)$ as
\begin{equation}\label{entropy def}
H(\Omega) = -\sum_{k=1}^{K}\frac{|\omega_k|}{N}\log\frac{|\omega_k|}{N}.
\end{equation}
Note that this functional form is often used for entropy in many other contexts different than information theory.\\
To find the relation between entropy and the Morisita index, we order our \textit{individuals} according to the box in which they are situated. We take $K$ to be the number of non-empty boxes. If we give these $K$ boxes indices $k$, then $|\omega_k|$ denotes the number of individuals in (non-empty) box $k$. Note that there is a slight subtlety here. Our boxes are what are called \textit{clusters} in the jargon of information theory (i.e. to which individuals are assigned). These should not be confused with our intuitive interpretation of `clusters': aggregations of individuals. Only in specific cases (dense aggregations or large boxes) such aggregation will be contained in a box.\\
\begin{figure}[!ht]
\centering
\includegraphics[scale=0.5]{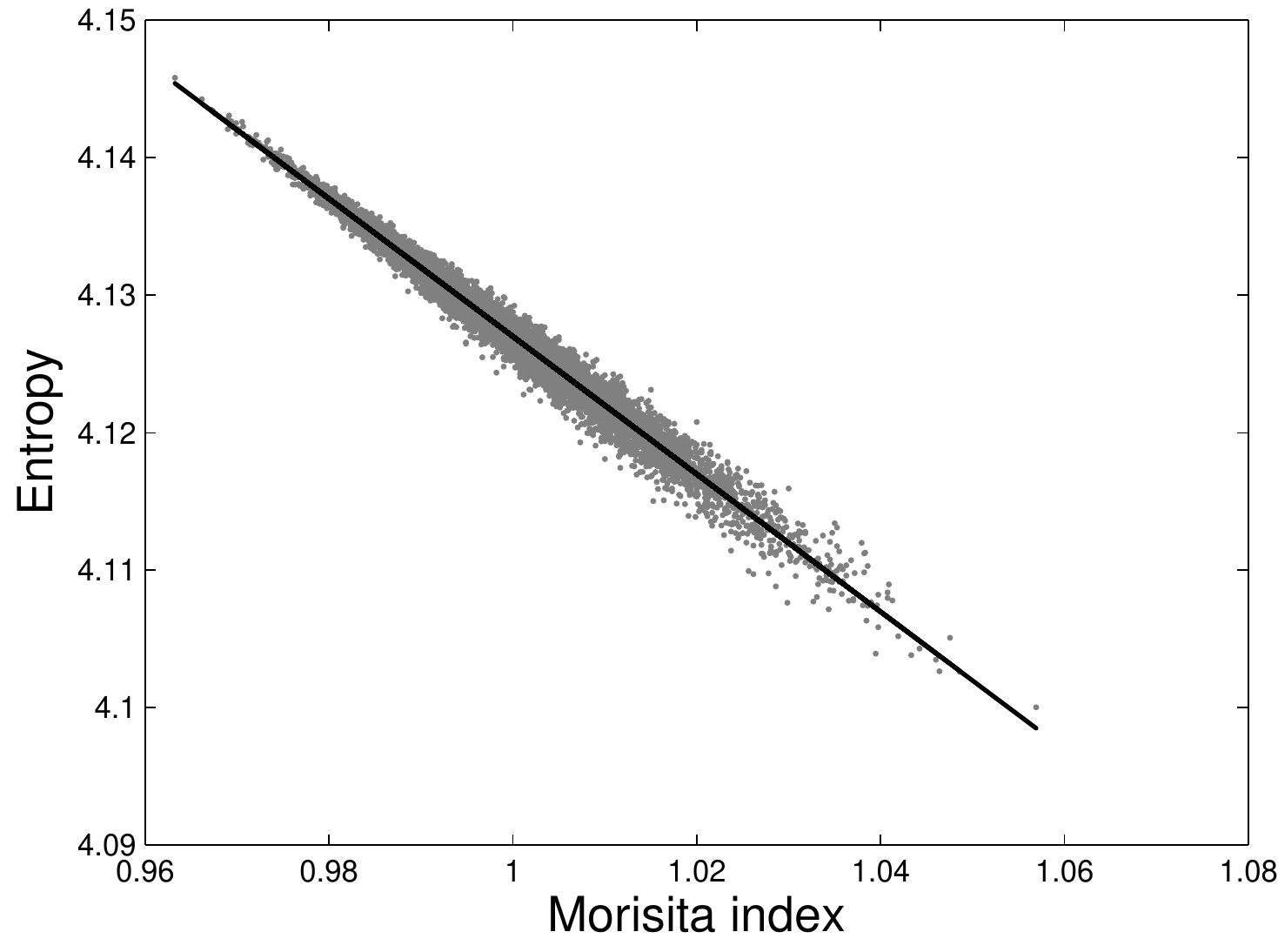}
\caption{Scatter plot of the entropy against the Morisita index for 10000 random configurations. A linear fit is added to the plot to emphasize the clear correlation between the two concepts.}
\label{Morisita Entropy fit}
\end{figure}

\noindent We investigate numerically the relation between entropy and the Morisita index. Each individual in a group of size $N=1000$ is assigned randomly to one of in total $\mathcal{M}=64$ boxes. Note that this value of $\mathcal{M}$ corresponds to the reference value in Table \ref{modpar}, which will also be used in the sequel. Using the obtained configuration we calculate the Morisita index (\ref{Morisita def}) and the entropy (\ref{entropy def}). This procedure is repeated 10000 times. For both quantities the exact position of an individual does not matter; it is only important to know in which box he is. Therefore we only need to assign a box (randomly). In Figure \ref{Morisita Entropy fit} we plot the entropy against the Morisita index for each realization. There is a clear correlation between the two, which we indicate by adding a linear fit of the data.\\
\\
Although entropy and Morisita are not the same concepts (and do not have the same interpretation), our numerical investigation supports the idea that they are similar in some sense. There is a clear (negative) correlation, which means that \textit{mutatis mutandis} we can deduce the same information from either of the two measures.

\subsection{Measured quantities: results}\label{sect: measured quant results}

\subsubsection{Polarization}\label{sect:results polarization}
\begin{figure}[!ht]
\begin{minipage}[t]{0.5\linewidth}
\centering
\includegraphics[scale=0.5]{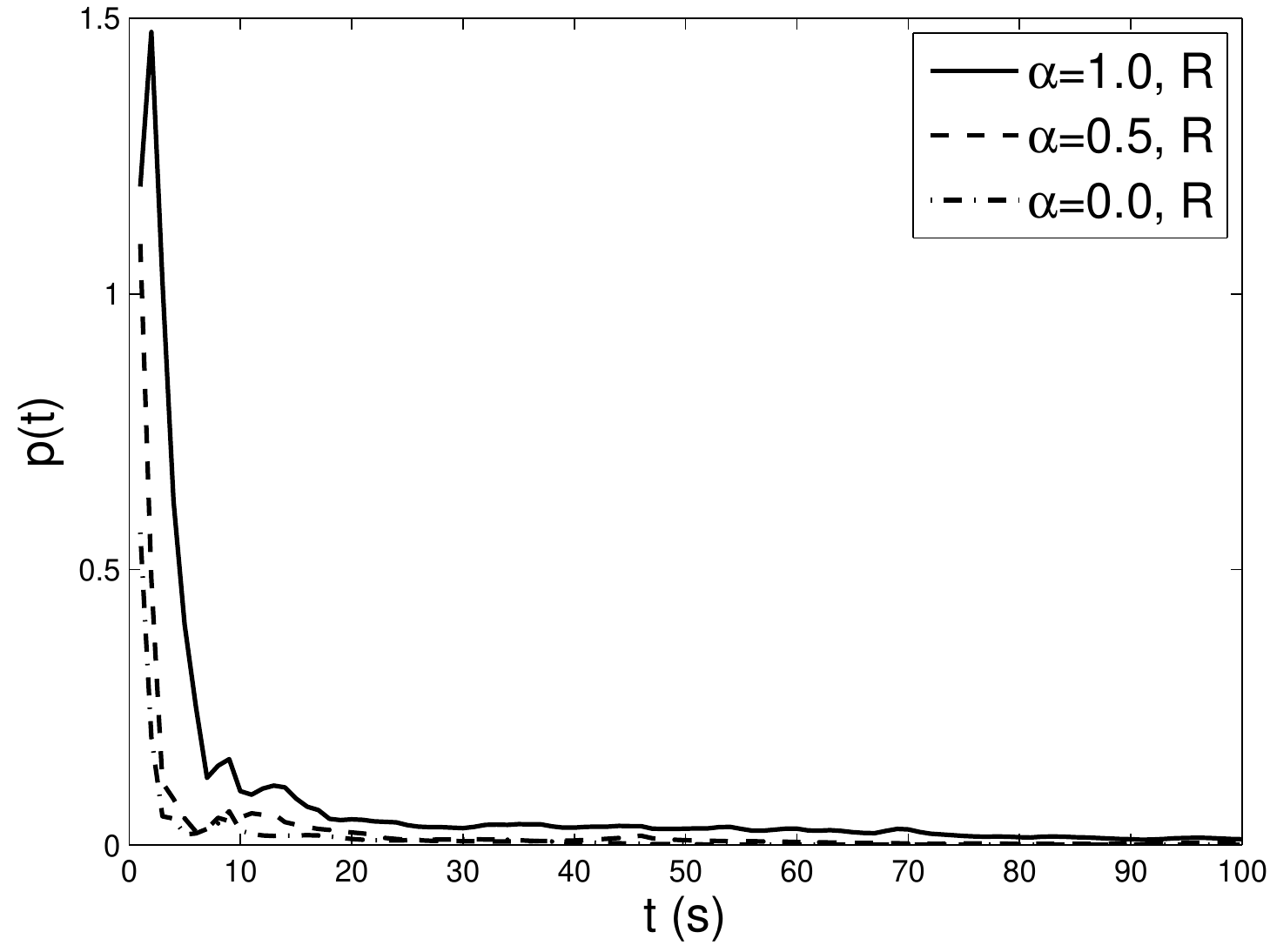}
\caption{The instantaneous polarization $p$ as a function of time. Results in the \underline{R case} for $N=100$, for several different values of $\alpha$.}
\label{pol t r}
\end{minipage}
\begin{minipage}[t]{0.5\linewidth}
\centering
\includegraphics[scale=0.5]{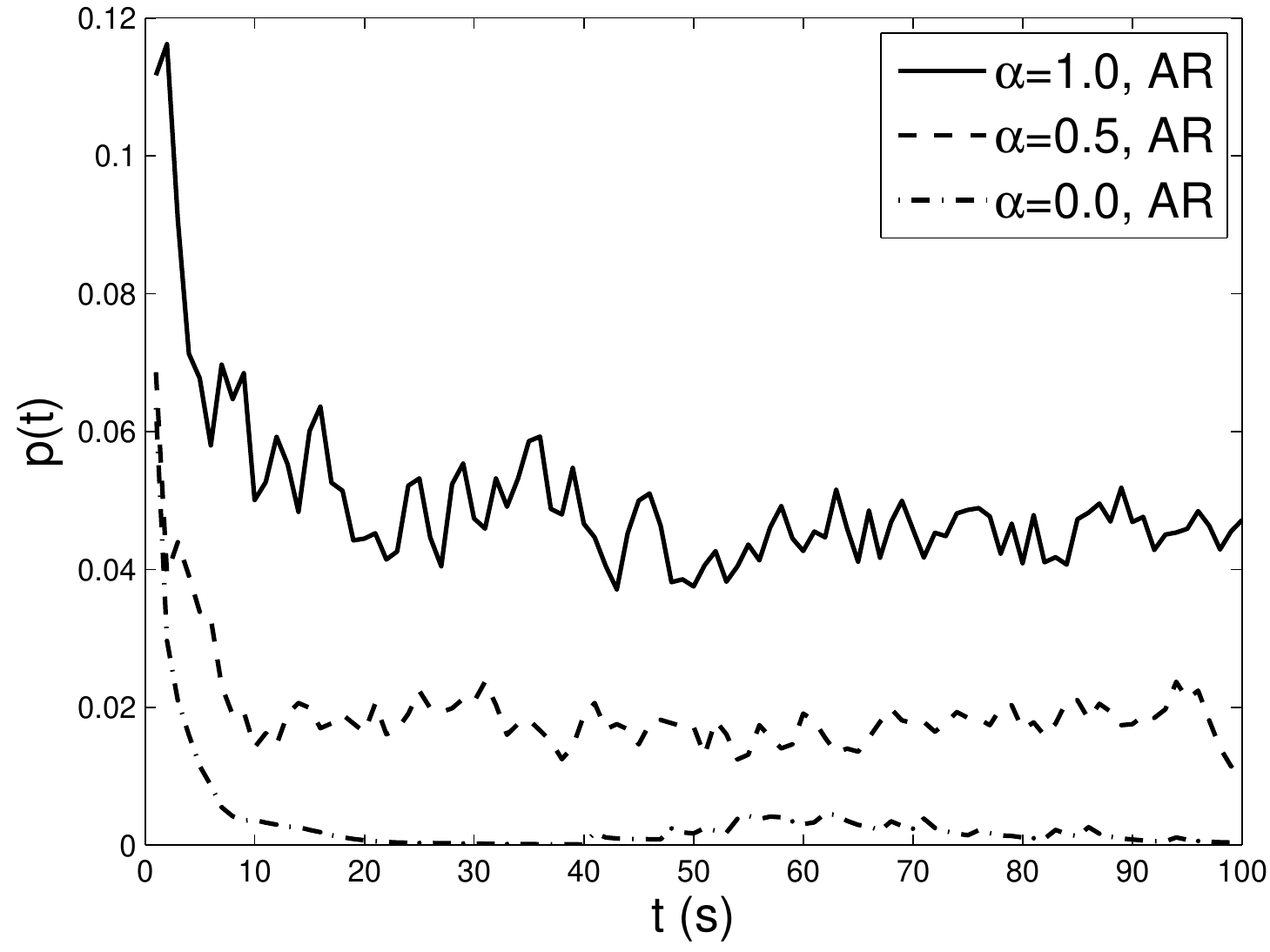}
\caption{The instantaneous polarization $p$ as a function of time. Results in the \underline{AR case} for $N=100$, for several different values of $\alpha$.}
\label{pol t ar}
\end{minipage}
\end{figure}

To show the kind of information we can deduce from the polarization index, we start by examining the evolution of $p(t)$ for $N=100$. The combined results are shown in Figures \ref{pol t r} and \ref{pol t ar}.\\

\noindent In each of the six cases we identify a relatively short period of time just after $t=0$, during which the polarization decreases rapidly. Afterwards, there is a state in which $p$ fluctuates around a certain average level; this state might be an equilibrium. For each $\alpha$ in the \underline{R case}, and for $\alpha=0.0$ in the \underline{AR case}, this average level is zero (we needed to zoom in in Figure \ref{pol t r} to verify that these curves really decay below the level as in Figure \ref{pol t ar}). The scenarios $\alpha=1.0$ and $\alpha=0.5$ in the \underline{AR case} are different in the sense that they do not decay to zero. Moreover, if one would zoom in, one would see that the oscillations in the \underline{R case} are less rapid and of smaller amplitude than in the \underline{AR case}. This holds especially for $\alpha=0.5$ and $\alpha=0.0$. Comparing R to AR requires some extra care, however. It is difficult to compare them in a fair way, because of their intrinsically different nature, and because of the (in)compatibility of the tested values for the interaction radii (cf. Table \ref{modpar}).\\
\\
Let us focus on the initial rapid decay of $p$: this suggests that the initial configuration is not a favourable state for the system to be in. An immediate relaxation takes place, implying spreading of the individuals in all possible directions until a more preferable situation is reached. Figures \ref{pos_r_00}--\ref{pos_ar_05} support this statement. Most clearly in Figures \ref{pos_ar_00} and \ref{pos_ar_05} the system evolved away from the initial configuration. The same statement is true for Figures \ref{pos_r_00}, \ref{pos_r_05} and \ref{pos_r_10}, although this is less evident. Once we realize however that the particles move in what seem to be \textit{six} horizontal rows, we must indeed conclude that the particles have deviated from the initial situation in which there were \textit{ten} rows. Figure \ref{pos_ar_10} is somewhat different, even though we recognize relaxation, as the initial occupation was only about one tenth of the corridor length. Spreading in a direction parallel to the mean direction of motion does not explain the peak in $p$ just after $t=0$ however. This is because fluctuations in the magnitude of the velocity (that is: the speed) do not affect the polarization if all individuals move in the same direction. The peak shows that there must have been some vertical displacement too.\\
\\
We remark that in the \underline{AR case} $p$ does not necessarily decay to zero, while it does (or at least: \textit{seems to do} up to fluctuations and noise) in the \underline{R case}. Regarding Figures \ref{pos_r_00}--\ref{pos_ar_10}, we can distinguish the graphs for $\alpha=1.0$ and $\alpha=0.5$ in the \underline{AR case} from the four others, since they do not seem to possess the degree of order and structure that the other graphs do have. Strikingly, these are exactly the cases in Figure \ref{pol t ar} where $p$ does not tend to zero. From this we conclude that there is a strong relation between the polarization tending to zero, and the preservation or favouring of patterns and organization in the system. The cases in which $p$ oscillates around a non-zero average in the long run, are exactly those in which the initial ordered configuration has disappeared after some time. Moreover, the fact that the polarization remains positive is an indicator that the configurations in Figures \ref{pos_ar_00} and \ref{pos_ar_05} are not stable.\\
\begin{figure}[!ht]
\begin{minipage}[t]{0.5\linewidth}
\centering
\includegraphics[scale=0.5]{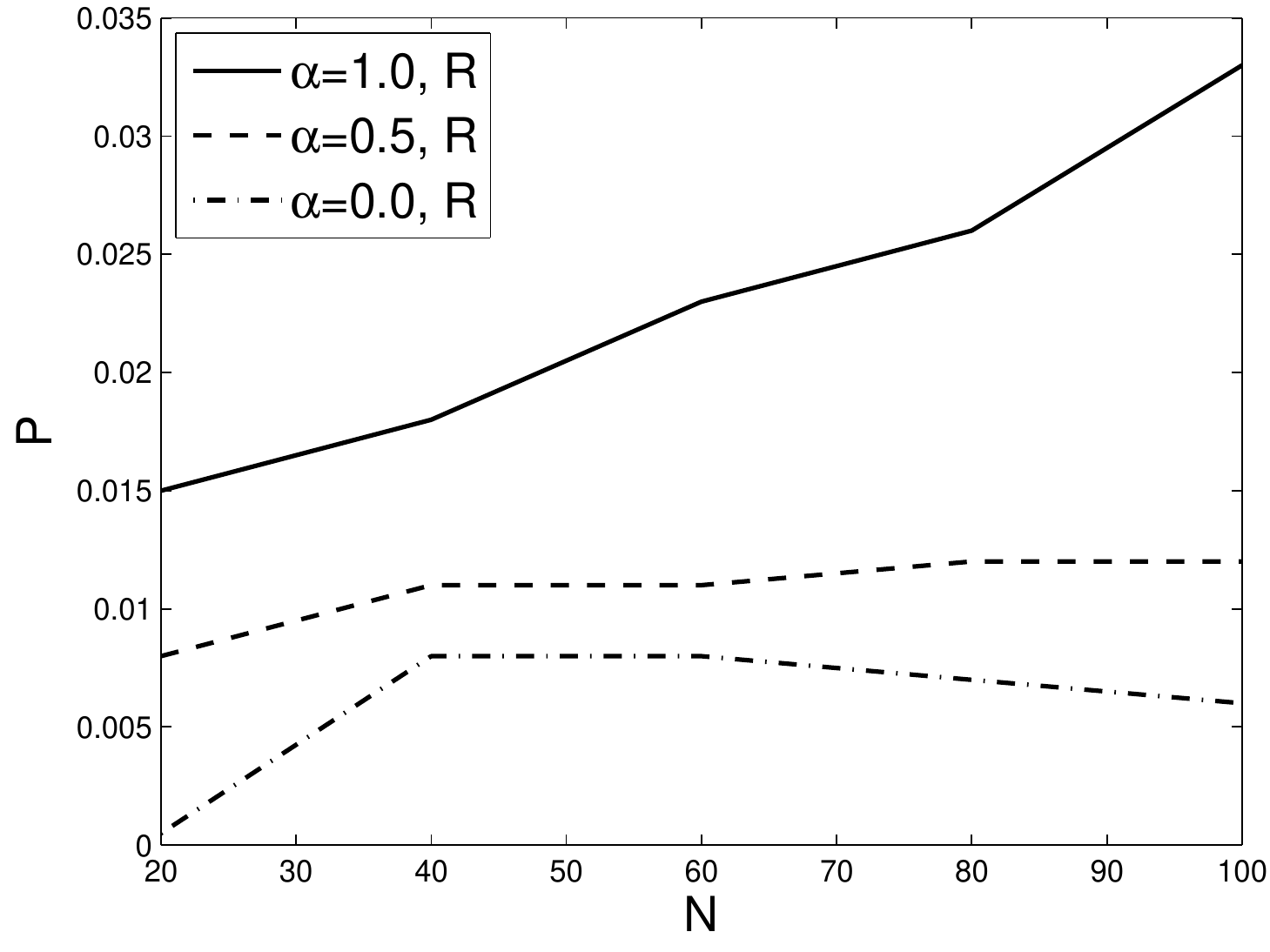}
\caption{The time average of the polarization $P$ as a function of the number of pedestrians $N$. Results in the \underline{R case}, for several different values of $\alpha$.}
\label{npr}
\end{minipage}
\begin{minipage}[t]{0.5\linewidth}
\centering
\includegraphics[scale=0.5]{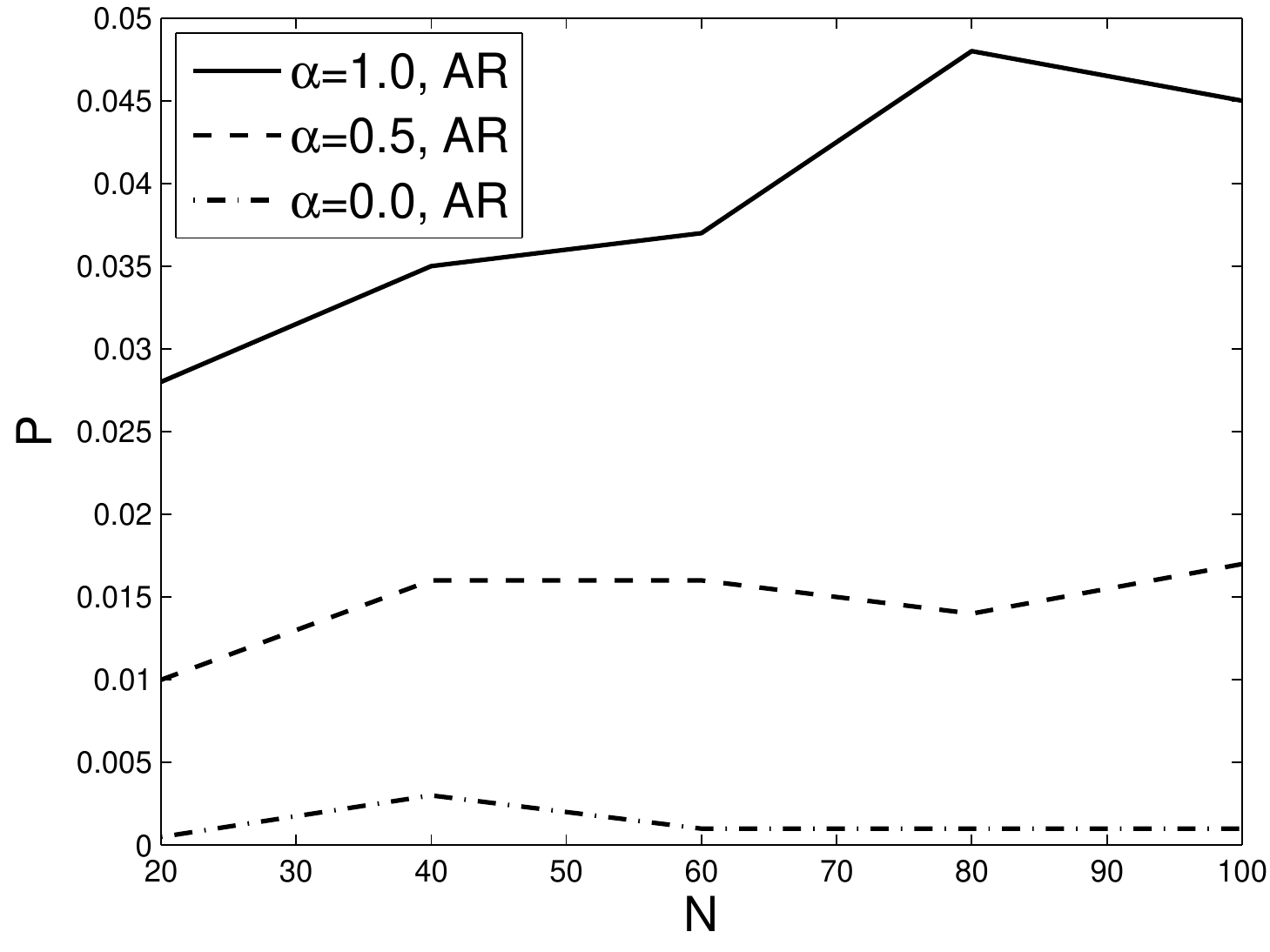}
\caption{The time average of the polarization $P$ as a function of the number of pedestrians $N$. Results in the \underline{AR case}, for several different values of $\alpha$.}
\label{npar}
\end{minipage}
\end{figure}

\noindent In Figures \ref{npr} and \ref{npar} the time average polarization index $P$ is shown. For each $N$, we recognize in $P$ the same ordering with respect to $\alpha$ as the ordering we have seen before in $p$ (just for $N=100$, cf. Figures \ref{pol t r} and \ref{pol t ar}). The ordering between $\alpha=0.5$ and $\alpha=0.0$ (\underline{R case}) is even much clearer when considering the time average.\\
Note that, comparing R and AR, the absolute differences in $P$ are much smaller than those in $p$. This is mainly because the $p$-curves in the \underline{AR case} have smaller range but larger domain where they are non-negligible.\\
\\
Up to now we only used initial conditions in which the individuals are positioned on a grid in only a part of the corridor. We wish to explore the influence of these initial conditions, by allowing randomness. More specifically, we sample the initial position of each individual from a random distribution on the whole corridor. The corresponding time averages of the polarization are given in Table \ref{pol random IC}, represented by the mean and standard deviation of 10 random runs. Both the \underline{R case} and the \underline{AR case} are considered.\\

\begin{table}[h]
\caption{Time average polarization after time $t = 100$s, for $N=60$ individuals and $\alpha=1.0$. The mean and standard deviation of 10 independent simulation runs are given. In each run the initial conditions were drawn from a random distribution, such that the initial positions are random over the whole corridor. We compare the outcome to the previously used initial conditions (i.e. on a lattice in a section of the corridor). The ratio: polarization for regular initial conditions divided by the average over 10 random runs is also given. Calculations were performed both for the \underline{R case} and the \underline{AR case}.\\}\label{pol random IC}
\centering
\begin{tabular}{r|rr}
\textrm{Polarization}&\textrm{R case}&\textrm{AR case}\\
\hline
Mean & 0.0387 & 0.0147\\
Standard deviation & 0.0019 & 0.0018\\
Regular initial conditions & 0.023 & 0.037\\
Ratio & 0.5941 &  2.5170\\
\end{tabular}
\end{table}
\noindent Unfortunately, these results do not provide an unambiguous conclusion about the effect of random initial conditions. In the \underline{R case} the mean polarization is bigger for random initial conditions than for regular ones. In the \underline{AR case} this is the other way around. An issue in interpreting the ratios in the bottom line of the table is that we are dividing relatively small numbers. However, this is something that needs to be stressed: the polarization remains relatively small, also for random initial conditions. We will see later (see e.g.~Table \ref{Morisita random IC}) that for the Morisita index the presented ratio does make sense, and that we can use them to draw conclusions about the effect of randomness.

\subsubsection{Projection of the pedestrian density on the $\vec{e}_x$ axis}\label{sec: result projection}
\begin{figure}[!ht]
\begin{minipage}[t]{0.5\linewidth}
\centering
\includegraphics[scale=0.5]{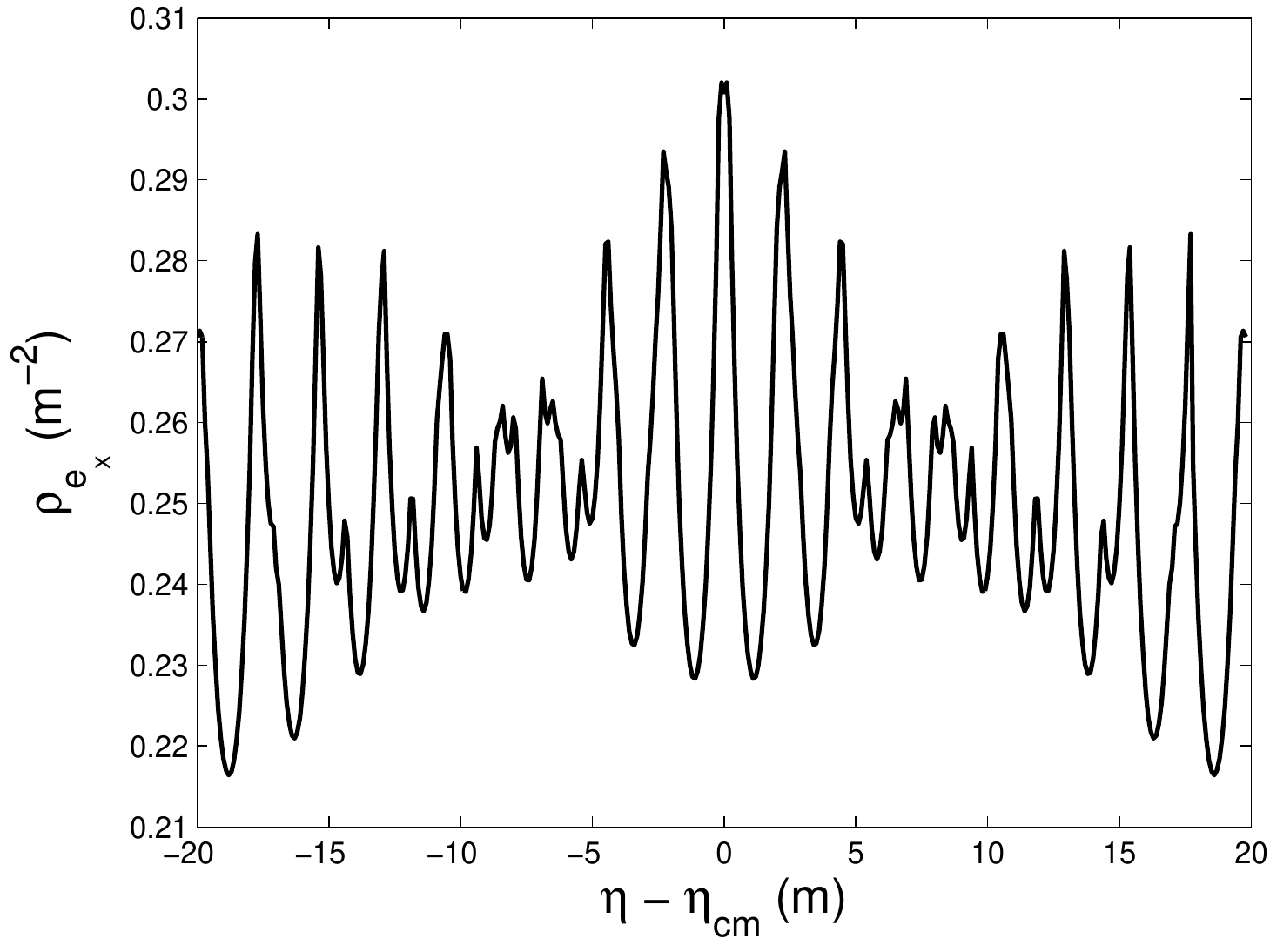}
\caption{The projection of the pedestrian density on the $\vec{e}_x$ axis $\rho_{\vec{e}_x}$ as a function of the spatial coordinate $\eta$, relative to the center of mass $\eta_{cm}$. Results in the \underline{R case} for $\alpha = 0.0$ at time $t = 100$s.}
\label{vdr_1_0}
\end{minipage}
\begin{minipage}[t]{0.5\linewidth}
\centering
\includegraphics[scale=0.5]{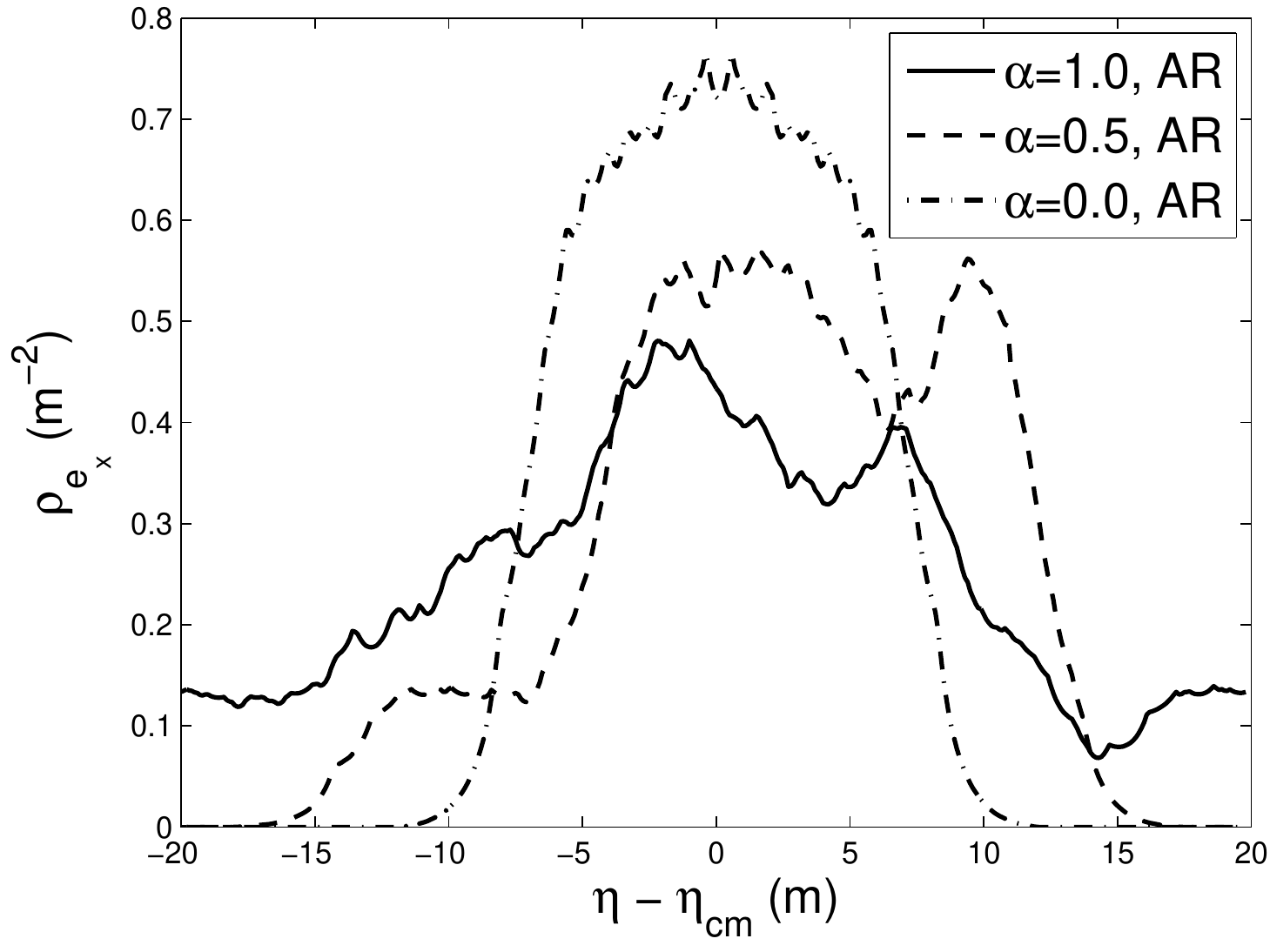}
\caption{The projection of the pedestrian density on the $\vec{e}_x$ axis $\rho_{e_x}$ as a function of the spatial coordinate $\eta$, relative to the center of mass $\eta_{cm}$. Results in the \underline{AR case} for several different values of $\alpha$ at time $t = 100$s.}
\label{vdar}
\end{minipage}
\end{figure}

Figures \ref{vdr_1_0} (\underline{R case}) and \ref{vdar} (\underline{AR case}) show the projected mollified density that was introduced in Section \ref{sect:Sim, projection}. The graphs show that in the isotropic case ($\alpha = 0.0$) the density profile is completely symmetric around the center of mass. Note that this is only the case if the initial conditions are symmetric. For $\alpha=1.0$ and $\alpha=0.5$ in the \underline{AR case} symmetry is no longer present. This is perfectly sane, since anisotropy (i.e. $\alpha>0$) was introduced to incorporate asymmetric interactions in our model: a pedestrian is more influenced by an other individual in front of him than by one behind him. The graphs corresponding to $\alpha=1.0$ and $\alpha=0.5$ in the \underline{R case} are not given. They possess the same oscillatory behaviour as Figure \ref{vdr_1_0}, but without being symmetric around the center of mass. They do not provide any further information or insight, and thus are omitted.\\

\noindent The highly oscillatory behaviour in Figure \ref{vdr_1_0} and the omitted graphs reflects the ordered, lattice-like, structures we already observed in Figures \ref{pos_r_00}--\ref{pos_r_10}. Following this reasoning, one would expect the same kind of oscillations also in the \underline{AR case} for $\alpha=0.0$ (cf.~the ordered pattern in Figure \ref{pos_ar_10}). The reason for not seeing this in Figure \ref{vdar} is simple: The average distance in $\vec{e}_x$ direction is smaller than the width of the support of our induced density. The total density therefore smoothens out the periodic structure of the individuals' positions. What we do see is the fact that in this situation the individuals do not occupy the whole corridor, but are confined to a certain section of it.\\
\\
What we observe is that including attraction in the interaction has some regularizing effect. For each of the three choices for $\alpha$, there is a core of high density around the center of mass, which forms the heart of our crowd. This core is present due to the initial condition that was concentrated on a part of the corridor. Without attraction the repulsive interactions would drive it apart. However, attraction is not able to completely diminish the effect of repulsion. Especially when $\alpha$ increases, the projected density of the core decreases, while there is a tail of mass just behind it (that is, in the graph on the left of the center). This effect can be explained by the fact that individuals in the anisotropic case are driven backwards if they are too close together (like in the high density core). There is no (sufficient) compensation driving the individual forward as this effect is decreased/switched off by increasing $\alpha$.\\
Apparently the mechanism that drives individuals to the back has more effect as $\alpha$ increases. In Figure \ref{vdar} we namely see that due the periodic boundaries the tail already reaches again the front of the core group. We expect that the Morisita index (see Section \ref{sect:results Morisita}) will provide extra support for our observations here; this index should be high when pedestrians concentrate more and more in a particular region.

\subsubsection{Morisita index}\label{sect:results Morisita}
\begin{figure}[!ht]
\begin{minipage}[t]{0.5\linewidth}
\centering
\includegraphics[scale=0.5]{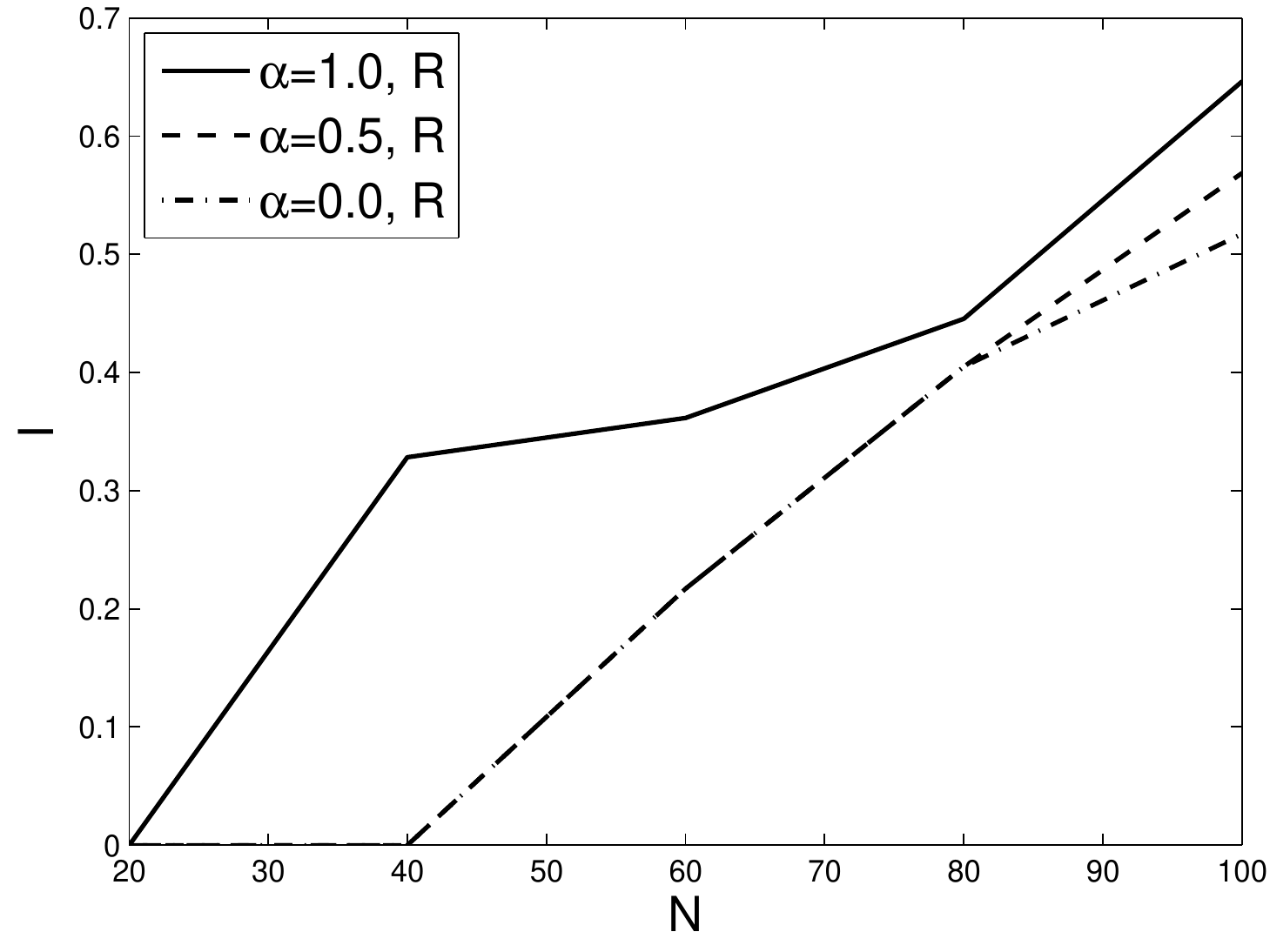}
\caption{The Morisita index $I$ as a function of the number of pedestrians $N$. Results in the \underline{R case}, for several different values of $\alpha$ at time $t = 100$s.}
\label{nir}
\end{minipage}
\begin{minipage}[t]{0.5\linewidth}
\centering
\includegraphics[scale=0.5]{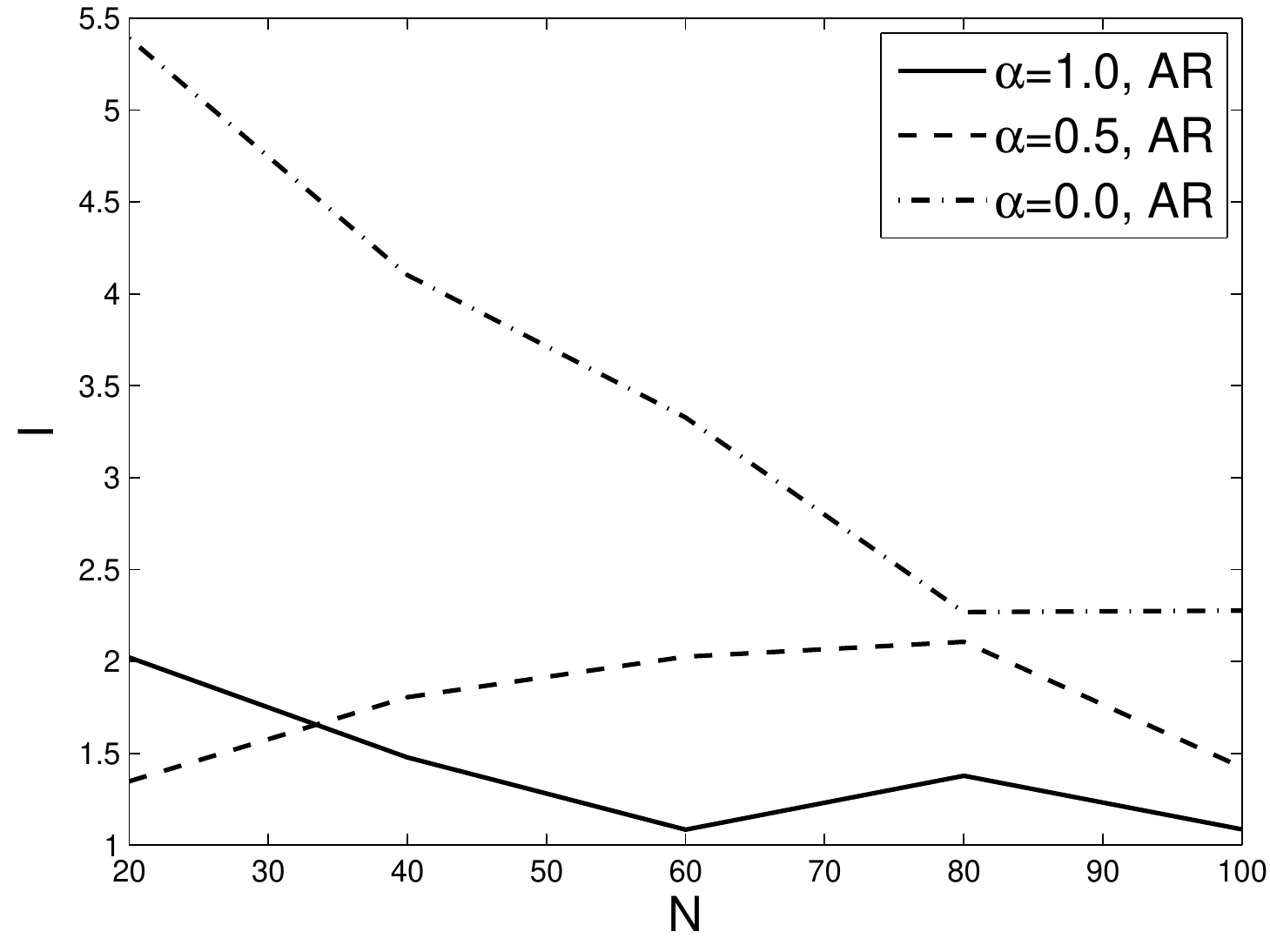}
\caption{The Morisita index $I$ as a function of the number of pedestrians $N$. Results in the \underline{AR case}, for several different values of $\alpha$ at time $t = 100$s.}
\label{niar}
\end{minipage}
\end{figure}

Figures \ref{nir} and \ref{niar} show the Morisita index as a function of the number of individuals $N$. The corridor is subdivided in $\mathcal{M}=64$ boxes.\\

\begin{figure}[!ht]
\centering
\includegraphics[scale=0.5]{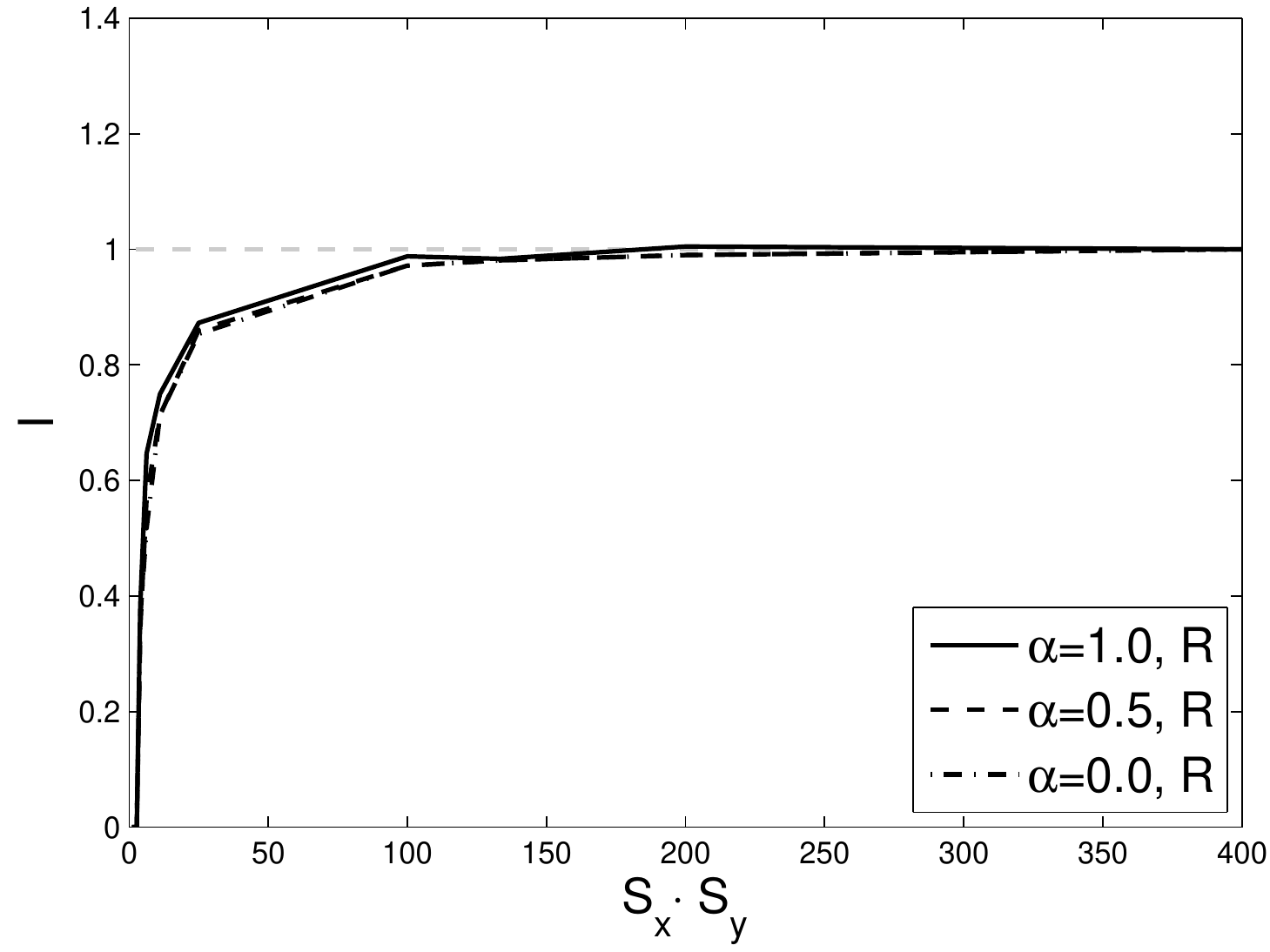}
\caption{The Morisita index $I$ as a function of the box size $S_x\cdot S_y$. Results in the \underline{R case}, for $N=100$, and for several different values of $\alpha$ at time $t = 100$s. The grey dashed line indicates the value 1. Note that in this figure we thus deviate from the (fixed) values for $\mathcal{M}$ and $S_x\cdot S_y$ given in Table \ref{modpar}.}
\label{size i r}
\end{figure}

\begin{figure}[!ht]
\centering
\includegraphics[scale=0.5]{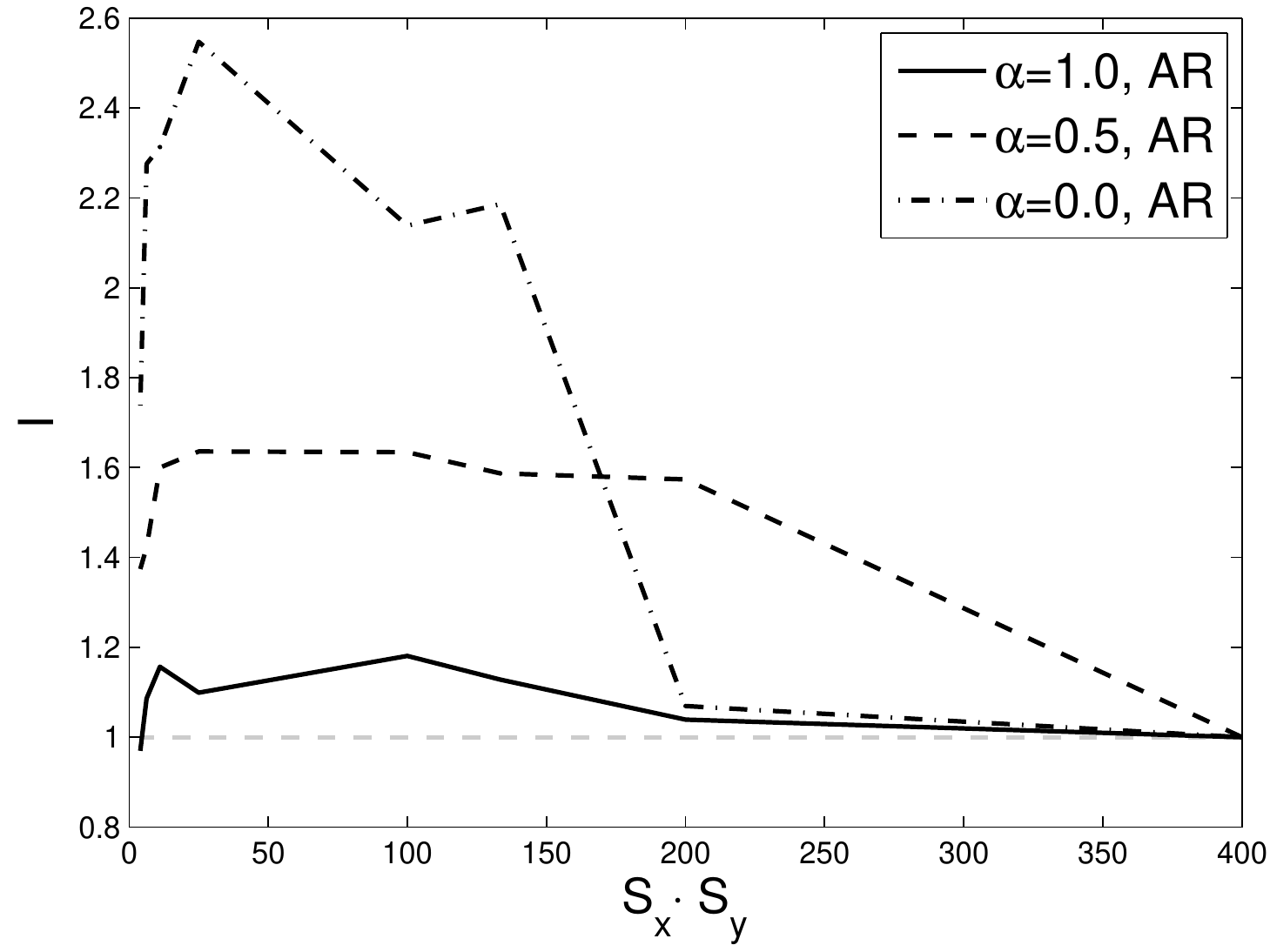}
\caption{The Morisita index $I$ as a function of the box size $S_x\cdot S_y$. Results in the \underline{AR case}, for $N=100$, and for several different values of $\alpha$ at time $t = 100$s. The grey dashed line indicates the value 1. In this figure we also deviate from the values for $\mathcal{M}$ and $S_x\cdot S_y$ given in Table \ref{modpar}.}
\label{size i ar}
\end{figure}

\noindent The increase in the Morisita index as $N$ increases (in the \underline{R case}) can be explained. For repulsive interactions, the individuals have a tendency to move as far apart as possible (if possible until they are a distance $R^R_r$ apart). This was illustrated by Figures \ref{pos_r_00}--\ref{pos_r_10}. As $N$ increases, they are however packed together more and more, thus leading to an increase in Morisita index. If we assume (to obtain an approximate result) that the individuals are distributed uniformly, then $n_i=N/\mathcal{M}$ for all $i$. Taking into consideration the ordered distribution in Figures \ref{pos_r_00}--\ref{pos_r_10}, this assumption is justifiable (see also below). It follows that $I_{\mathcal{M}}=(N-\mathcal{M})/(N-1)=1-(\mathcal{M}-1)/(N-1)\rightarrow 1$ as $N\rightarrow\infty$ for fixed $\mathcal{M}$. Moreover, this implies that $I_{\mathcal{M}}$ tends to its limit value from below. This matches with the increase of the curves in the \underline{R case} (Figure \ref{nir}), and we conjecture that the Morisita index will tend to 1 for $N$ increasing beyond $N=100$.\\
\\
There is actually more support for the assumption that the individuals in the \underline{R case} are distributed uniformly. This support is provided by Figure \ref{size i r}, in which (for $N=100$) we show the Morisita index as a function of the box size. In \cite{Morisita59} an overview is given of the information that can be derived from such graph. Note that the Morisita index equals 1 if the number of boxes $\mathcal{M}=1$. This corresponds to the largest possible box size. The value 1 is also indicated in the graph by the dashed grey line. In Figure \ref{size i ar} we see that for each of the three values of $\alpha$ the Morisita index approaches the value 1 from below as the box size increases. This particular trend in the graph corresponds to a uniform distribution of particles (cf. Figure 1 in \cite{Morisita59}).

\noindent In the \underline{AR case} (Figure \ref{niar}), we observe that the Morisita index is \textit{not} monotonic in $\alpha$ for $N$ smaller than $N\approx 35$ (this is where the graphs for $\alpha=1.0$ and $\alpha=0.5$ intersect). This point of intersection is hard to explain and requires further investigation. Moreover, it is hard to draw any conclusion about the precise dependence of the Morisita index on $N$ or $\alpha$ in this case. One could argue that increasing $\alpha$ corresponds to lower Morisita index. At least this is the case if one compares $\alpha=0.0$ to $\alpha>0$.\\
\\
We now consider the influence of varying the box size in the \underline{AR case}. The trend, as shown in Figure \ref{size i ar}, is different from the \underline{R case} (Figure \ref{size i r}). The most importance difference is that the value 1 is approached \textit{from above} for increasing box size. This case is described in Figure 1 of \cite{Morisita59} as `contagious distribution' with `clump(s)'. In our terminology this would correspond to clustering. We already observed that in the \underline{AR case} the individuals do not occupy the whole corridor (cf. Section \ref{sec: result projection}).\\

\noindent The shape of the graph and the way of approximating 1 from above, is linked in \cite{Morisita59} to the size of clumps (clusters) and the distribution of particles therein. We will not go into detail here, mostly since one can debate about the question to which class in Figure 1 of \cite{Morisita59} the graphs in Figure \ref{size i ar} should be assigned.\\

\begin{table}[h]
\caption{Morisita index after time $t = 100$s, for $N=60$ individuals and $\alpha=1.0$. The mean and standard deviation of 10 independent simulation runs are given. In each run the initial conditions were drawn from a random distribution, such that the initial positions are random over the whole corridor. We compare the outcome to the previously used initial conditions (i.e. on a lattice in a section of the corridor). The ratio: Morisita index for regular initial conditions divided by the average over 10 random runs is also given. Calculations were performed both for the \underline{R case} and the \underline{AR case}.\\}\label{Morisita random IC}
\centering
\begin{tabular}{r|rr}
\textrm{Morisita index}&\textrm{R case}&\textrm{AR case}\\
\hline
Mean & 0.1555 & 0.6979\\
Standard deviation & 0.0639 & 0.1079\\
Regular initial conditions & 0.3616 & 1.0847\\
Ratio & 2.3256 & 1.5544\\
\end{tabular}
\end{table}
\noindent To investigate the influence of the initial conditions, we also calculate the Morisita index corresponding to random initial conditions. More details are given in Table \ref{Morisita random IC}. Both in the \underline{R case} and in the \underline{AR case} the mean Morisita index over 10 random runs is significantly lower than the Morisita index for regular initial conditions. The difference is even much larger than the standard deviation.\\
However, there is consistency, in the sense that in both cases (R and AR) the values are lower.

\section{Conclusions and outlook}\label{sec:conclusion}
The behaviour of a crowd of pedestrians inside a corridor, in which the individuals interact via an anisotropic way, can be distinguished clearly from the case in which pedestrians interact in a completely isotropic way. In particular, we observe the following differences compared to the isotropic case:
\begin{enumerate}
\item the polarization index increases with increasing anisotropy (i.e. increasing $\alpha$);
\item the projected density along the $\vec{e}_x$ axis shows a symmetric profile around the center of mass for the isotropic case. However, increasing anisotropy implies loss of symmetry;
\item the Morisita index, as a measure of clustering, depends clearly on the anisotropy. It increases (with increasing anisotropy) in the \underline{R case} and, roughly speaking, decreases in the \underline{AR case}.
\item In case of repulsive interactions, the crowd tends to fill the whole corridor. If attraction is included, the group stays compact. Increasing $\alpha$ however seems to diminish this kind of \textit{social cohesion} as individuals do not look behind.
\end{enumerate}
As a result of our study, many new questions arose. Future research should be concentrated on the following three directions:
\begin{itemize}
\item Most obviously: what is the effect of a further increase of the number of pedestrians? Do the observed relations between the measured quantities and the number of pedestrians still hold? How do the observed limiting values of the polarization for large $t$ depend on $N$? What about the limit $N\rightarrow\infty$ in this case?\\
    Can we gain more insight in the strange issues of the \underline{AR case} (cf.~Figure \ref{niar}: intersecting curves for $\alpha=1.0$ and $\alpha=0.5$) by further measurements of the Morisita index for increasing $N$? E.g.~can we extrapolate information for $N>100$ back into the interval $[0,100]$? Some preliminary comments in this direction are given in \ref{Appendix}.\\
    If $N$ increases, the natural thing to do is to consider the discrete-to-continuum limit (i.e.~construct educated procedures to derive mean-field limit equations). Does such limit exist, can we derive it, and can we compare the effect of anisotropy in the limit to the observations of the current work?
\item How much does the large time behaviour of the crowd depend on the initial conditions? The initial distribution of pedestrians in this paper is not a realistic situation. People starting to enter a corridor are in real life never distributed in a crystalline structure. However, for an escape situation (for example in the case of fire) it seems reasonable to assume that a group of people starts, being clustered, at one side of a corridor. Therefore, as an extension of this research, we propose to use as initial distribution a more realistic configuration in which people are placed at one side of the corridor, with their positions (slightly) perturbed from the grid points. Averaging over a large collection of such perturbed initial distributions, will lead to effective results. Are these averaged results comparable to the ones presented in this paper? In other words: is averaging the results basically the same as removing the fluctuations from the initial conditions?\\
    In the paper we have included some preliminary results (in the ends of Sections \ref{sect:results polarization} and \ref{sect:results Morisita}) in this direction. There we took the other extreme: random initial conditions over the whole corridor.
\item What happens if we try to make our model more realistic: e.g. change the shape of the domain $\Omega$, or allow variation in the direction and magnitude of individuals' desired velocity? Including more sophisticated active parts in the boundary (doors) or impermeable objects within the domain, automatically leads to questions about the efficiency of the flow (such issues are also addressed e.g.~in \cite{Guo11, Guy}). Which geometry leads to the fastest evacuation? First steps in this direction have been made in \cite{EversMuntean}.
\end{itemize}
The issues addressed in this paper show that anisotropy related to perception has nontrivial effects on the global dynamics of a crowd. Certainly, these effects cannot be neglected. More work, both numerically and analytically, is needed to extend and formalize our results.

\ack The authors thank C. Storm, F. Toschi,  F. van de Ven, H. Wyss (all with TU Eindhoven) and R. Fetec\u au (Simon Fraser Univ.~Canada) for a series of fruitful discussions on the dynamics of self-propelled particles with anisotropic motion potentials. They are also grateful to J. Lega (Univ.~of Arizona USA) for sharing her thoughts. Moreover, they thank the anonymous referees for their comments and suggestions for improvements. JE kindly acknowledges the financial support of the Netherlands Organisation for Scientific Research (NWO), Graduate Programme 2010.

\appendix
\section{Towards a bigger number of particles}\label{Appendix}
\begin{figure}[!ht]
\begin{minipage}[t]{0.5\linewidth}
\centering
\includegraphics[scale=0.5]{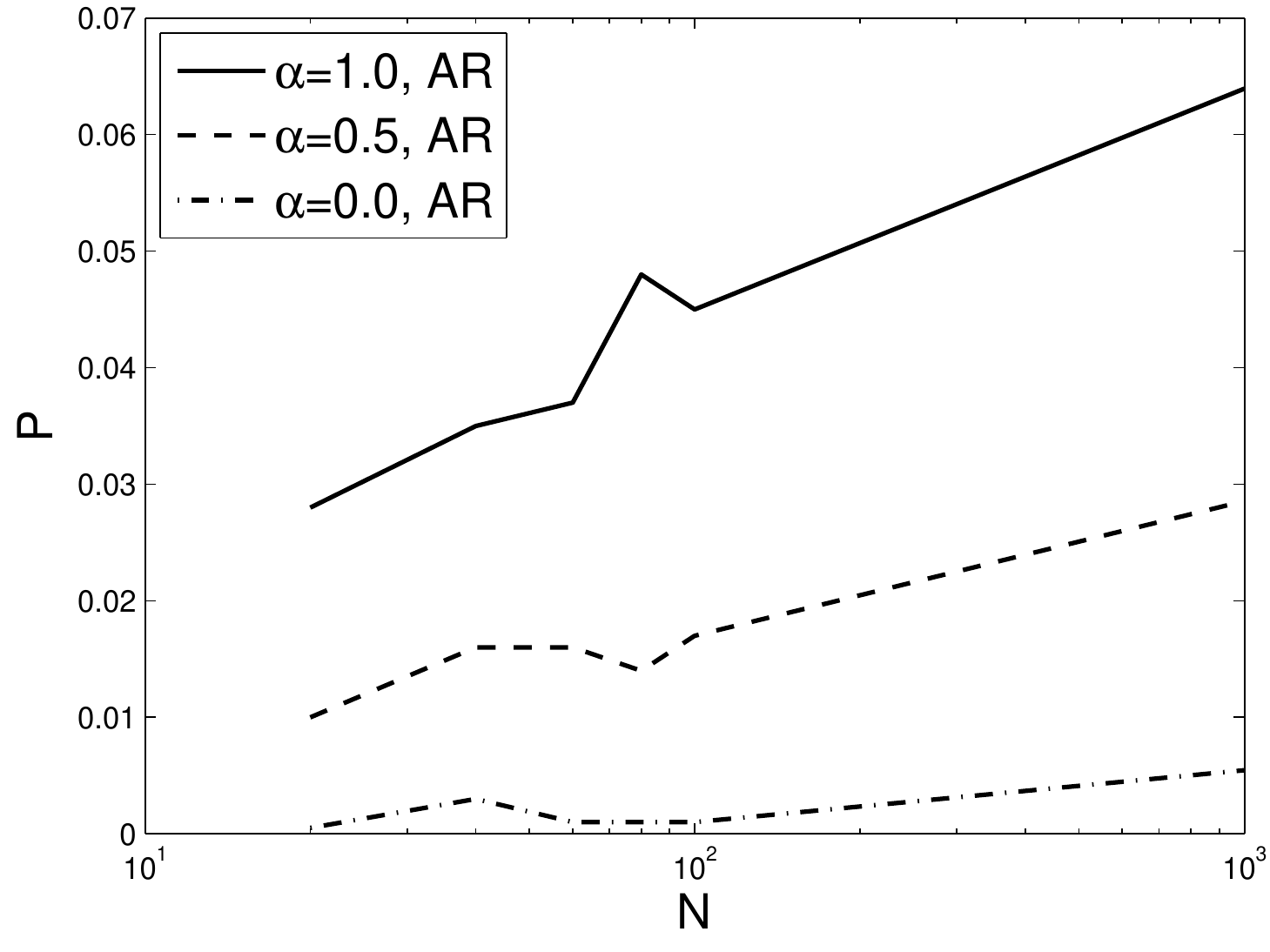}
\caption{The time average (over an interval of 100 s) of the polarization $P$ as a function of the number of pedestrians $N$. Results in the \underline{AR case}, for several different values of $\alpha$. This is an extension of Figure \ref{npar} including the value for $N=1000$ with logarithmic scaling on the horizontal axis.}
\label{p1000}
\end{minipage}
\begin{minipage}[t]{0.5\linewidth}
\centering
\includegraphics[scale=0.5]{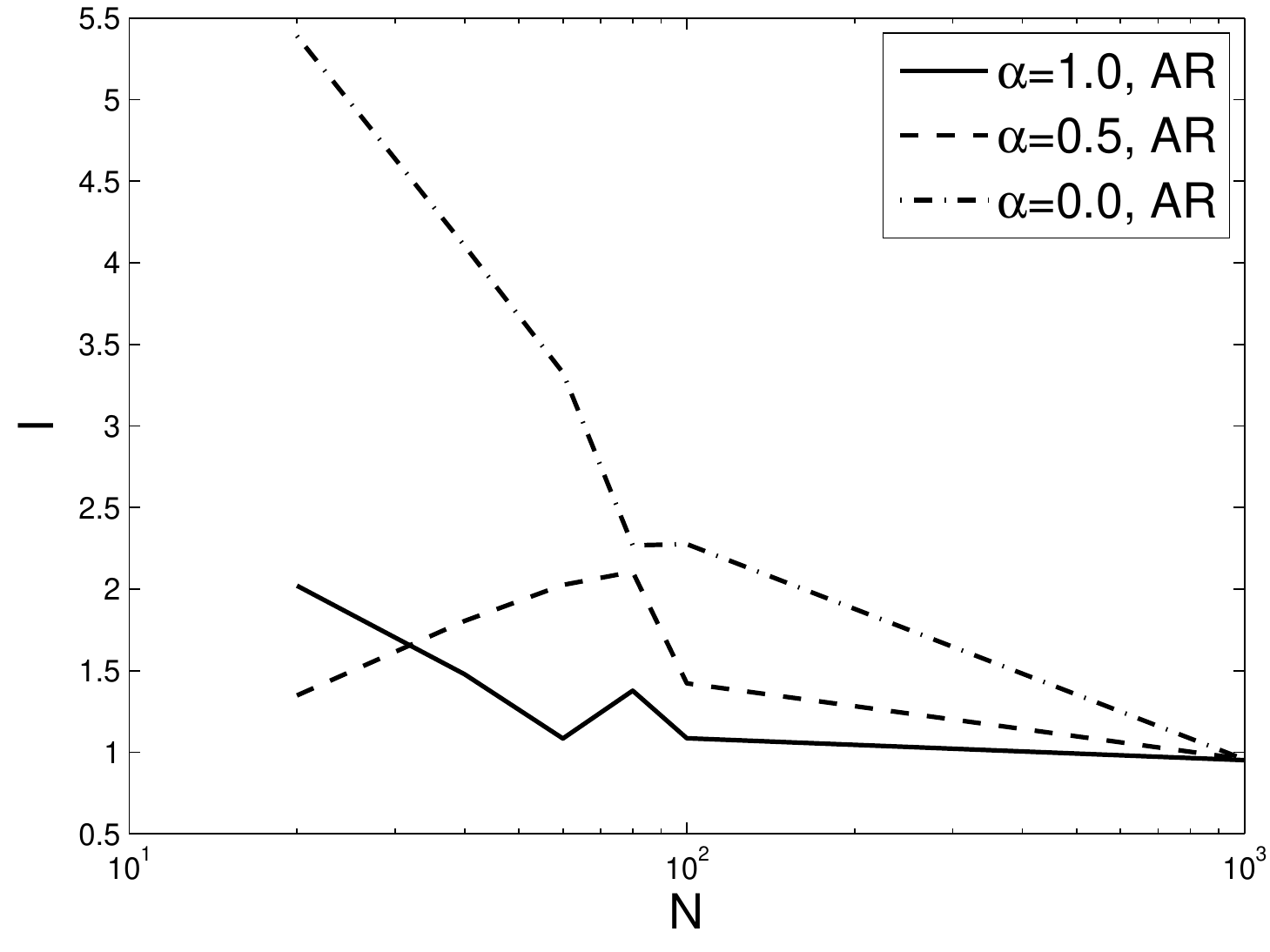}
\caption{The Morisita index $I$ as a function of the number of pedestrians $N$. Results in the \underline{AR case}, for several different values of $\alpha$ at time $t = 100$s. This is an extension of Figure \ref{niar} including the value for $N=1000$ with logarithmic scaling on the horizontal axis.}
\label{i 1000}
\end{minipage}
\end{figure}

Let us start by saying that simulation of large numbers of individuals is beyond the scope of this paper. Our current implementation is inadequate for simulating system sizes one is used to in molecular dynamics. In our perspective this paper aims primarily at getting insight about what features to expect. A second stage (and follow-up paper) is to optimize the implementation and increase the system size.\\
\\
Looking ahead, we provide here some preliminary results for $N=1000$ in the \underline{AR case}. In Figures \ref{p1000} and \ref{i 1000} we show the time-averaged polarization and the Morisita index, respectively, as a function of $N$. Compared to Figures \ref{npar} and \ref{niar}, the graphs have been continued by incorporation of the values at $N=1000$. Note the logarithmic scale of the horizontal axes.\\

\noindent Note moreover that we have extended Figures \ref{npar} and \ref{niar} by only one data point each. The linear interpolation between the values at $N=100$ and $N=1000$ is therefore probably not very meaningful. What we are interested in, is the general trend.\\
\\
For the polarization, the ordering as a function of $\alpha$ remains as we observed it. What requires more investigation is the increasing trend: that is, increasing $P$ for increasing $N$. In Figure \ref{npar} the graphs seem to stop growing as $N$ goes towards $100$. An issue here might be that the time interval of 100 s is simply too short for larger $N$.\\
\\
In the Morisita plots we see a downward trend. We see that the curve for $\alpha=1.0$ remains beneath the other two, also for $N=1000$. We can thus regard the absence of ordering of the curves for $N$ smaller than $N\approx 35$ as an exception.\\
A remark needs to be made about the fact that, at $N=1000$, the Morisita index for $\alpha=0.0$ is smaller than for $\alpha=0.5$. The values are $I= 0.9515$ for $\alpha=1.0$, $I=0.9583$ for $\alpha=0.5$ and $I=0.9566$ for $\alpha=0.0$. The ordering of the curves therefore seems to be lost here also, be it that the difference is small compared to the magnitude of $I$.

\end{document}